\documentclass[acmsmall]{acmart}
\AtBeginDocument{%
  }

\acmJournal{TOSEM}

\usepackage[utf8]{inputenc}
\usepackage{array}
\usepackage{wrapfig}
\usepackage{multirow}
\usepackage{tabularx}
\usepackage{caption}
\usepackage{subcaption}
\usepackage{booktabs}
\usepackage{setspace}
\usepackage{tikz}

\usepackage{amssymb}
\usepackage{enumitem}
\usepackage{multirow}
\usepackage{longtable}
\usepackage{changepage}
\usepackage{hyperref}
\hypersetup{
    colorlinks=true,
    linkcolor=black,
    filecolor=black,      
    urlcolor=black,
    citecolor=black,
}
\usepackage{graphicx}
\usepackage{longtable}
\usepackage{adjustbox}

\begin{document}

\title{Accessibility Recommendations for Designing Better Mobile Application User Interfaces for Seniors}

\author{Shavindra Wickramathilaka}
\email{shavindra.wickramathilaka@monash.edu}
\affiliation{%
  \institution{Faculty of Information Technology, Monash University}
  \city{Melbourne}
  \state{Victoria}
  \country{Australia}
}

\author{John Grundy}
\affiliation{%
  \institution{Faculty of Information Technology, Monash University}
  \city{Melbourne}
  \country{Australia}}
\email{john.grundy@monash.edu}

\author{Kashumi Madampe}
\affiliation{%
  \institution{Faculty of Information Technology, Monash University}
  \city{Melbourne}
  \country{Australia}}
 \email{kashumi.madampe@monash.edu}

\author{Omar Haggag}
\affiliation{%
 \  \institution{Faculty of Information Technology, Monash University}
  \city{Melbourne}
  \country{Australia}}
\email{omar.haggag@monash.edu}

\newcommand{\toolname}[0]{}
\renewcommand{\toolname}[0]{{\textit{AdaptForge}}}

\begin{abstract}
Seniors represent a growing user base for mobile applications; however, many apps fail to adequately address their accessibility challenges and usability preferences. To investigate this issue, we conducted an exploratory focus group study with 16 senior participants, from which we derived an initial set of user personas highlighting key accessibility and personalisation barriers. These personas informed the development of a model-driven engineering toolset, which was used to generate adaptive mobile app prototypes tailored to seniors' needs. We then conducted a second focus group study with 22 seniors to evaluate these prototypes and validate our findings. Based on insights from both studies, we developed a refined set of personas and a series of accessibility and personalisation recommendations grounded in empirical data, prior research, accessibility standards, and developer resources, aimed at supporting software practitioners in designing more inclusive mobile applications.
\end{abstract}

\ccsdesc[500]{Software and its engineering}
\ccsdesc[500]{Software and its engineering~Designing software}
\ccsdesc[500]{Human-centered computing~Accessibility}
\ccsdesc[500]{Human-centered computing~Empirical studies in accessibility}

\keywords{Accessibility Guidelines, Seniors, Older Adults, User Interfaces, Mobile Applications, Personalisation}


\maketitle

\section{Introduction}\label{sec1:introduction}

The World Health Organization (WHO) projects that by 2030, one-sixth of the global population will be aged over 60 years \cite{who2024}. Despite being a rapidly growing demographic segment, seniors continue to face numerous accessibility barriers in software user interfaces, particularly in the context of mobile application usage \cite{bossini2014,Paez2019, connor2017,wickramathilaka2025}. This situation is partly attributable to software practitioners who often overlook seniors as active users during the design and development of digital systems \cite{johnson2017designing}.
However, it is essential to note that these barriers are not typically introduced with intent. In our prior research, we found that software developers often express empathy toward the accessibility challenges faced by seniors \cite{wickramathilaka2025,vu2022better}. Many practitioners have older parents, relatives, or friends who experience similar difficulties when using mobile applications, and in some cases, developers themselves may also begin to encounter such barriers. This growing awareness underscores the need to integrate senior-inclusive design practices into mainstream software development.

Nonetheless, awareness alone does not necessarily translate into action. A key reason is the inherent complexity of addressing accessibility for seniors in applications. Age-related impairments can vary significantly \cite{czaja2007}. For example, some users may experience limitations in vision, hearing, mobility, or cognition, sometimes in combination. Designing for such a diverse user base is both time-consuming and resource-intensive \cite{vu2022better,liu2022curated,wickramathilaka2025}. Furthermore, seniors differ significantly in terms of background, language proficiency, personal preferences, and technological experience, which adds an additional layer of complexity due to the need for personalisation \cite{czaja2007, yigitbas2020}.

One of the most effective and comprehensive tools available to software practitioners is the use of established accessibility guidelines, such as the Web Content Accessibility Guidelines (WCAG) \cite{wcag2.2} and ISO 9241-171:2008 - \textit{Ergonomics of human-system interaction, Part 171: Guidance on software accessibility}, to support the development of universally accessible applications \cite{wickramathilaka2023}. We initially aimed to make use of these standards by specifically considering WCAG as the foundational framework behind a novel low-code method to address the age-related accessibility and personalisation needs of seniors \cite{wickramathilaka2023}. However, we realised that the senior-specific requirements identified in our research could not be fully addressed through these standards alone, as several critical gaps became evident. 

To address the gaps in existing accessibility guidelines, we first sought to deepen our understanding of age-related accessibility needs through a dual-phase focus group study involving 38 senior participants. To identify solutions to the range of accessibility issues uncovered during this user study, we began by examining two of the most widely recognised accessibility standards: WCAG 2.2 \cite{wcag2.2} and ISO 9241-171 \cite{iso9241}, both of which consider accessibility from the perspectives of developers and end users. To address additional gaps, we reviewed guidelines and heuristics proposed by researchers who conducted empirical studies directly with senior participants \cite{redeap, Shamsujjoha2025, inostroza2016, watkins2014, morey2019}, offering insights that are often more closely aligned with the needs of end users. Lastly, we incorporated the developer’s perspective into this mix by examining platform-specific resources, including Apple’s \cite{iosDevGuidelines} and Google’s \cite{androidDevGuidelines} accessibility guidelines.

The following research questions guided this process:

\begin{enumerate}[label={RQ (\arabic*)}]  
    \item \textit{What accessibility barriers do seniors face when using mobile applications? } 
    \item \textit{How can developers mitigate these accessibility barriers when designing mobile applications?}
\end{enumerate}

Additionally, we recognised that applying static, one-size-fits-all design guidelines alone would be insufficient to address a critical class of user needs, namely, the personalisation needs of seniors. Therefore, we extended our user studies to investigate how these personalisation needs shape accessibility requirements, leading to the following research question:

\begin{enumerate}[resume,label={RQ (\arabic*)}]
    \item \textit{What are the barriers and enablers for seniors in personalising their mobile application experience?}
\end{enumerate}

\noindent The key contributions of this paper are as follows:  
\begin{itemize}
    \item Identification of the key accessibility needs of diverse seniors based on empirical evidence via a dual-phased focus group study with senior mobile app end users;
    \item Illustration of these key accessibility and personalised needs of diverse senior mobile app end users through a set of senior personas; 
    \item A set of accessibility recommendations for developing more personalised apps for seniors, consolidated through both empirical evidence and an analysis of existing accessibility standards and guidelines; and  
    \item A range of insights into seniors' perceptions of potential run-time app adaptations to meet their individual accessibility and app personalisation needs. 
\end{itemize}

The structure of the rest of this paper is as follows: Section \ref{sec2:motivation} presents the motivation behind this study and a summary of related works, followed by an explanation of the research methodology in Section \ref{sec3:methodology}. Section \ref{sec4:problems and solutions} provides an in-depth analysis of the qualitative data gathered from our focus groups, discussing accessibility problems and their solutions. Section \ref{sec5:limitations} presents a discussion of limitations and future works. Finally, we conclude the paper with a summary in Section \ref{sec6:summary}.

\section{Motivation and Related Work}\label{sec2:motivation}

\begin{figure}
\includegraphics[width=\textwidth]{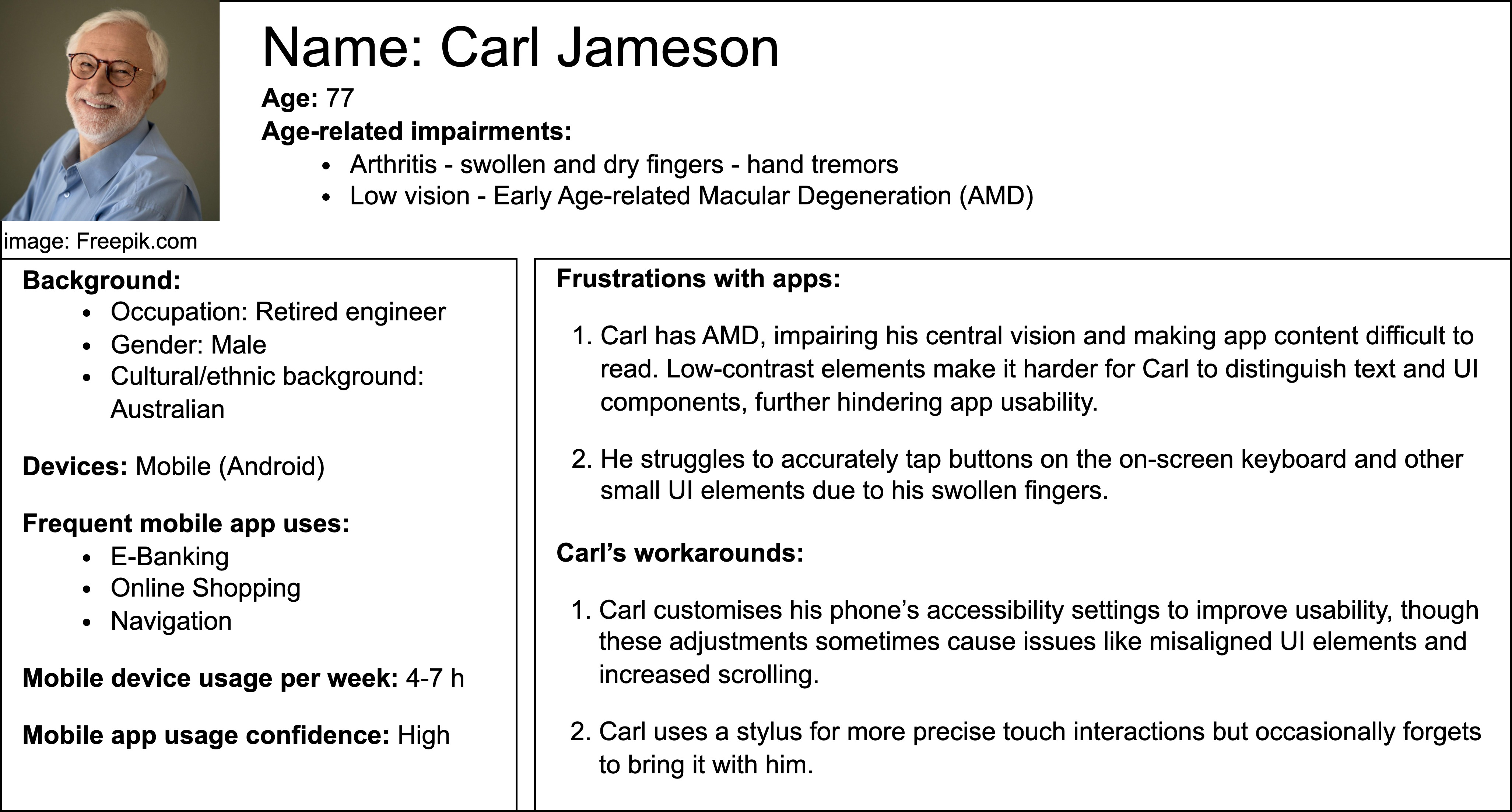}
\caption{A persona representing a fictional user named Carl, who is confident in using technology due to his professional background, yet still encounters significant accessibility barriers in mobile applications due to vision and mobility impairments.}\label{persona example}
\Description{A persona of a senior named Carl who is 77 years of age. He has both arthritis and low vision, and this has caused him to have frustrations: First is his Age-Related Macular Degeneration condition; his central vision is impaired, and it is making app content difficult to read. Low-contrast elements make it harder for Carl to distinguish text and UI components, further hindering app usability. Another issue is that he struggles to accurately tap buttons on the on-screen keyboard and other small UI elements due to his swollen fingers. Carl customises his phone’s accessibility settings to improve usability, though these adjustments sometimes cause issues like misaligned UI elements and increased scrolling. He also uses a stylus for more precise touch interactions but occasionally forgets to bring it with him.}
\end{figure}

\subsection{Seniors' Motivation}

We wanted to better understand the diverse usability and accessibility needs of senior mobile app end users to identify key personalisation strategies that could be employed to improve the apps. To this end, we conducted a dual-phase focus group study. Consider the persona of Carl, illustrated in Figure~\ref{persona example}, which was derived from the findings of that study. Carl represents a senior individual who has attempted to customise font size to improve readability, only to encounter misaligned user interface (UI) elements as a consequence. Additionally, he experiences reduced hand dexterity, making interacting with the on-screen keyboard difficult. To address this, he uses a third-party stylus as a personal workaround.

Carl’s experience with accessibility barriers was found to be common among seniors in our study. A recurring theme was the widespread use of personal workarounds to manage and mitigate accessibility challenges. For example, one participant reported keeping a physical notebook to manually record navigational paths within mobile applications. These findings suggest that many seniors have adapted to their circumstances through considerable effort and creativity. However, the responsibility of managing accessibility barriers should not fall on the users. Instead, researchers and software practitioners should take responsibility for developing and integrating evidence-based solutions into the software development process to address these issues proactively.

\subsection{Developers' Motivation}

To understand the origins of accessibility barriers for seniors, we also examined this issue from the perspective of software developers, as the root cause of many of these challenges lies in their design and development decisions. Our motivation to explore this problem from a developer-centric viewpoint stems from our own background as software practitioners, enabling us to relate closely to our target audience. As part of our study, we interviewed software developers with varying levels of experience to understand their needs and requirements for a development tool (\toolname) aimed at building adaptive and accessible apps for seniors \cite{wickramathilaka2025}. Our findings revealed that the majority of participants (14 out of 18) expressed a deep personal empathy toward the challenges seniors face when using the apps they develop \cite{wickramathilaka2025}.

Another key finding from our developer interview study was that when asked about the strategies that developers would employ to develop UIs apps for seniors, 2/3 of the developers responded that they would consider a universally accessible UI based on existing industry standards \cite{wickramathilaka2025}. The next popular strategy (from 11 out of 18 participants) is to design apps based on the guidelines and compliance requirements native to the platforms they are developing the apps for \cite{wickramathilaka2025}. The first strategy relies on platform agnostic accessibility standards such as the extensive collection of standards defined and iteratively improved by the World Wide Web Consortium (W3C) such as Web Content Accessibility Guidelines (WCAG) \cite{wcag2.2}, the Accessible Rich Internet Applications (WAI-ARIA) suite \cite{wai-aria}, the User Agent Accessibility Guidelines (UAAG) \cite{uaag}, and the Authoring Tool Accessibility Guidelines (ATAG) \cite{atag}. Another widely recognised standard is the Ergonomics of Human-System Interaction - Part 171: Guidance on Software Accessibility (ISO 9241-171:2008) \cite{iso9241}, developed by the International Organization for Standardization (ISO). For the second strategy, developers generally make use of platform-specific resources such as Apple's developer iOS/iPadOS accessibility guidelines \cite{iosDevGuidelines} and Google's Android developer guidelines \cite{androidDevGuidelines}.

\subsection{Diverse Senior App End User Personas}

Despite the presence of empathetic developers and the availability of established accessibility guidelines, mobile applications continue to present significant accessibility barriers for seniors \cite{Paez2019}. This persistent issue, previously identified by other researchers, prompted us to engage directly with a local senior community to investigate the real-world accessibility challenges they encounter. Through a series of focus groups involving 38 senior participants, we consistently identified critical accessibility issues affecting their day-to-day app usage.

In this paper, we have sought to represent the voices of our participants and their key frustrations and challenges through a series of fictional personas (Figures~\ref{persona example}, \ref{judy_persona}, \ref{kathryn_persona}, \ref{liam_persona}, \ref{usha_persona}, \ref{ava_persona}, \ref{stefan_persona}, \ref{dorothy_persona}). These personas are grounded in the lived experiences of seniors and reflect the diversity and severity of the accessibility barriers they face. As software practitioners ourselves, we approached this research with empathy and a sustained commitment to addressing senior users' accessibility needs. Yet, throughout this multi-year study, we arrived at a sobering realisation: Despite our technical expertise, familiarity with accessibility guidelines, and genuine intent to design inclusively, the existing standards and platform-specific developer resources proved insufficient to fully address the issues uncovered in our user study. This experience has underscored for us the urgent need to expand and refine current accessibility frameworks to better support the real-world needs of senior users.

To further contextualise this issue, we refer to the experience of a developer who participated in our prior interview study \cite{wickramathilaka2025}. \textit{Darrow}, a fictional persona based on this participant, was a highly experienced developer tasked with building an accessible Digital Traveller Declaration app for New Zealand. His experience working with accessibility standards is summarised in the following quote:

\begin{quote}
    \textit{"We tried to meet accessibility requirements for mobile [app]. However, the bigger challenge was actually figuring out what that [accessibility] standard was that we had to meet, and the consulting company that assessed whether we met those needs or not, had a different view of what we had [done]. [...] And as a result of that project, the Department of Internal Affairs [in Government] has updated their guidelines for meeting accessibility needs."}
\end{quote}

This example illustrates the complexities and ambiguities developers often face when trying to comply with accessibility requirements. Therefore, the limitations of current developer resources must be considered from multiple perspectives. Our work draws on feedback from senior users, empirical findings from other studies involving diverse senior cohorts, and analysis of existing standards and platform-specific development guidelines to understand both end-user and developer perspectives.
By integrating these insights, we propose a set of accessibility design recommendations that aim to address the needs of both seniors and developers in the context of mobile app development. While our recommendations may not be exhaustive, we believe their practical grounding and empirical basis will make them valuable for developers building industry applications and for standard-makers working to close this critical societal accessibility gap.

\subsection{Related Work}\label{sec7:related work}

We consulted two primary sources to guide our exploration into accessibility guidelines for the senior user community: (1) established accessibility guidelines and standards; and (2) prior research studies that propose design heuristics or recommendations specifically for mobile applications targeting seniors.
Within the first category, the Web Content Accessibility Guidelines (WCAG) 2.2 \cite{wcag2.2} emerged as the most prominent, comprehensive, and up-to-date set of accessibility guidelines applicable to the design of mobile applications for seniors. Another influential standard is ISO 9241-171:2008 - \textit{Ergonomics of human-system interaction, Part 171: Guidance on software accessibility} \cite{iso9241}. Although it is relatively dated, having been last reviewed and confirmed in 2018, our experience suggests it remains a valid and valuable source for accessible software design.

In addition to these general standards, we also examined platform-specific accessibility guidelines: Apple's Developer Accessibility Guidelines \cite{iosDevGuidelines} and Google's Android Developer Accessibility Guidelines \cite{androidDevGuidelines}. The decision to include these sources was influenced by insights from a prior qualitative study we conducted with software developers \cite{wickramathilaka2025}, which revealed that a key developer priority when designing apps for seniors is adherence to platform-native accessibility recommendations and configurations. Accordingly, it was logical for us to include these platform guidelines in our analysis to derive relevant design insights.
The second source category comprised design guidelines and heuristics proposed in prior empirical studies. We sought to evaluate whether the themes identified in these studies align with the Accessibility Problem Dimensions (APD) uncovered in our own research with seniors. Of 23 studies identified, the predominant focus areas were cognitive aspects (e.g., memory limitations, cognitive overload, and need for clear guidance) and visual aspects (e.g., font legibility and colour contrast). More specifically, we categorised the existing recommendations as follows:

\begin{enumerate}[label=APD\arabic*:]
    \item Enhancing text readability: \cite{chai2017, redeap, kurniawan2008, Shamsujjoha2025, watkins2014, lee2018, antonelli2018, morey2019}.
    \item Improving colour contrast: \cite{redeap, salman2018, kurniawan2008, Shamsujjoha2025, harte2017, watkins2014, morey2019}.
    \item Improving touch interaction target accessibility: \cite{Shamsujjoha2025, mi014, harte2017, almeida2015, barros2014, antonelli2018, morey2019, leitao2012}.
    \item Addressing limited screen space: \cite{redeap, inostroza2016, salman2018, harte2017, barros2014, antonelli2018, leitao2012}.
    \item Error handling, jargon, and language
        \begin{enumerate}[label=\roman*.]
            \item Error handling: \cite{redeap, inostroza2016, salman2018, Shamsujjoha2025, watkins2014, barbosa2015}.
            \item Use of accessible language and avoiding jargon: \cite{redeap, Shamsujjoha2025, harte2017, watkins2014, almeida2015, barros2014, morey2019}.
        \end{enumerate}
    \item Mitigating fear of making mistakes: \cite{harte2018, morey2019, holzl2013, almeida2015, conte2019, philips2018, barbosa2015, holden2020, watkins2014, harte2017, Shamsujjoha2025, kurniawan2008, salman2018, inostroza2016}.
    \item Supporting app navigation for seniors: \cite{redeap, Shamsujjoha2025, mi014, almeida2015, barros2014, lee2018, morey2019}.
    \item Inclusion of audio modalities (input/output): \cite{Shamsujjoha2025, mi014, barros2014, redeap}.
\end{enumerate}

As we present our proposed guidelines, we elaborate on how these guidelines and heuristics informed our design rationale and the accessibility recommendations proposed for seniors. In addition, the literature synthesis matrix we provided in Figure \ref{literature_matrix} provides an overarching view of how in-alignment or in-conflict our recommendations are compared to existing standards, guidelines, and studies.

\section{Our Approach}\label{sec3:methodology}

\subsection{Research Methodology Overview}

During the conception of our study, we recognised that seniors face accessibility barriers in software applications, particularly on mobile devices. This understanding stemmed from both the researchers' personal experiences and existing literature. However, to deepen our understanding and refine our research questions, we sought to explore this phenomenon in a real-world context. To achieve this, we initially conducted a pilot focus group study with a senior community in Australia. Insights from this pilot informed a subsequent focus group with different participants from the same community, marking the exploratory phase of our study.

\begin{figure}
\includegraphics[width=\textwidth]{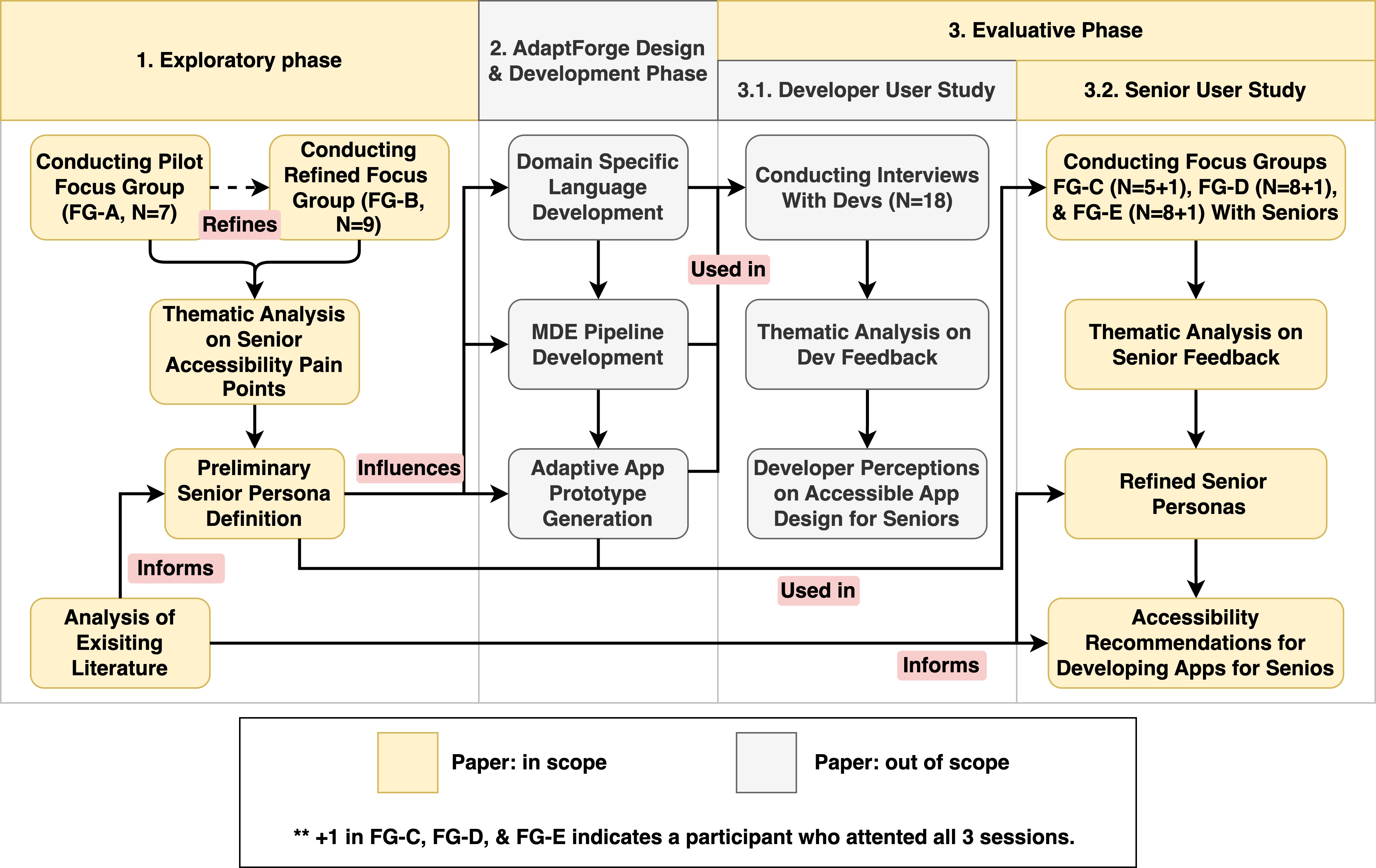}
\caption{An overview of the focus group study design}\label{overview}
\Description{The figure presents an overarching view of the study methodology. The first phase is exploratory, comprising two focus group studies. Insights from this phase informed Phase 2, which involved the development of adaptive mobile app prototypes. Phase 3 encompassed the evaluation stage, consisting of two user studies: one with senior end-users and another with software developers. It is important to note that this paper focuses solely on the exploratory phase and the senior user study conducted during the evaluation phase.}
\end{figure}

Findings from this exploratory phase guided the creation of design artefacts, shaping the conceptualisation and implementation of \toolname, a tool designed to help developers generate adapted app UI instances at design time in its current prototype form. These adapted app instances were then used as examples in a final set of focus groups during the evaluative phase to validate our findings, reassess our approach, and inform future enhancements.
An overview of our research methodology is presented in Figure \ref{overview}.

\subsection{Exploratory Phase}

At the outset of our study, we made two key assumptions: (1) seniors face accessibility barriers in app UIs due to age-related needs, and (2) individuals have diverse requirements, meaning that one senior’s UI needs may differ from another’s. To evaluate these assumptions and enhance our understanding of the phenomenon, we chose focus groups as our research method. This approach enabled interactive discussions among senior participants, allowing them to share their experiences with app UIs and providing us with a broader understanding of their accessibility challenges. Additionally, we aimed to identify both common requirement patterns among participants and instances where individual needs diverged. Finally, we sought to collate the findings from our exploratory focus groups and our analysis of existing literature and accessibility standards to generate a preliminary set of personas to inform design decisions in subsequent stages of our study.

\subsubsection{\textbf{Exploratory Focus Groups: Design}}

First, we obtained approval from the Monash Human Subject Ethics Committee (MUHREC Project ID: 32963) and then conducted a pilot focus group to explore the problem space. To facilitate discussion among seniors, we employed a presentation-based approach, displaying various app UI examples and gathering participant feedback. For instance, to assess the reading experience on mobile apps, we presented a comparative visual scenario: one image featured illegible text requiring users to zoom in for readability, while the other displayed easily readable text with a larger font and a bold header. Similarly, other questions addressed key accessibility aspects, including comprehension of app content (considering technical jargon, language differences, and cultural factors), navigation challenges, interaction with input methods such as touchscreens and virtual keyboards, the use of voice input and screen readers, and engagement with device or app accessibility settings.

Thereafter, we leveraged our experience from the pilot study to refine our discussion facilitation method, improving our questions, examples, and demographic data collection process. These enhancements enabled us to conduct a more structured and insightful focus group. We subjected the data from both focus groups to thematic analysis, yielding key insights into real-world age-related accessibility challenges. This understanding served as the ground truth for the subsequent design and development phase of our project.
We explain how we recruited our participants in the exploratory stage and their demographic information later in Section \ref{exploratory_participants}.

\subsubsection{\textbf{Exploratory Focus Groups: Data Collection}}

Both focus groups were held in person and conducted by the authors, each running for approximately an hour. We audio-recorded the sessions. The FG-B focus group recording was transcribed manually by an author, while FG-A was transcribed using the Otter.ai transcription tool.

\subsubsection{\textbf{Exploratory Focus Groups: Data Analysis}}

We manually cleaned both transcripts and subjected the data to a thematic analysis process. Nvivo was used as a tool for qualitative analysis. Initially, we coded the data to identify a broad range of enablers and barriers that seniors encounter in their day-to-day lives when using apps. These codes were then refined through subsequent iterations.

\subsubsection{\textbf{Exploratory Focus Groups: Participants}} \label{exploratory_participants}

For the pilot focus group FG-A, we recruited 7 senior participants from a University of the Third Age (U3A)\footnote{\href{https://u3aaustralia.org.au/}{U3A Australia}} chapter in Australia. For the subsequent ‘refined' focus group FG-B, we recruited another 9 participants in a different U3A chapter. Each participant was gifted with a 30 AUD gift voucher from a local supermarket as a token of appreciation.

The demographic information of our exploratory focus group study participants is provided in Table \ref{tab:exploratory_demographic_data}. A total of \(N_{AB} = 16\) seniors participated in the exploratory phase, with \(N_A = 7\) seniors in the pilot study and \(N_B = 9\) seniors in the subsequent refined focus group. The mean age of participants was 73.9 years, while the median age was 74 years. The majority of participants were female (75\%, \(n = 12\)) and from an Australian cultural background (68.8\%, \(n = 11\)).  
In addition, all participants reported using spectacles, while a minority (25\%, \(n = 4\)) also reported using hearing aids. Almost all participants (93.8\%, \(n = 15\)) owned a smartphone, and a majority also reported owning a tablet device (\(n = 12\), 75\%) and a personal computer (desktop or laptop) (56.3\%, \(n = 9\)).

During the pilot study, we collected data on participants' app usage types and the approximate time they spent weekly on apps using a show-of-hands method during the focus group session. As a result, we were unable to assign the identified data to individual pilot study participants. However, we revised the demographic data collection form to include these two questions, allowing us to map this information to individual participants in the subsequent focus group.  

These questions enabled us to identify the types of apps frequently used by seniors. The most popular app domains among participants were online banking (62.5\%, \(n = 10\)), online shopping (50\%, \(n = 8\)), navigation (62.5\%, \(n = 10\)), social media (75\%, \(n = 12\)), video calls (62.5\%, \(n = 10\)), video streaming (56.3\%, \(n = 9\)), e-books (56.3\%, \(n = 9\)), and audiobooks (50\%, \(n = 8\)).
Participants also reported that the majority (75\%, \(n = 12\)) spend four or more hours per week using software applications. Finally, we assessed their confidence in IT-related skills (usage of apps and devices) to determine the general IT literacy and familiarity with technology usage within the population sample. Additionally, we aimed to examine whether varying levels of IT usage confidence influenced the personalisation preferences of our senior participants. A majority (68.8\%, \(n = 11\)) of participants reported that they have average or above levels of confidence in their IT usage-related skills.

\begin{table}
    \centering
    \scriptsize
    \renewcommand{\arraystretch}{1.3} 
    \resizebox{\textwidth}{!}{
    \begin{tabular}{|l|l|l|>{\raggedright\arraybackslash}p{1.9cm}|>{\raggedright\arraybackslash}p{2cm}|>{\raggedright\arraybackslash}p{2cm}|p{1.6cm}|>{\raggedright\arraybackslash}p{3cm}|p{1.5cm}|}   
        \hline
        \textbf{ID} & \textbf{Age} & \textbf{Gender} & \textbf{Culture} & \textbf{Usage of Aids} & \textbf{Device Usage} & \textbf{App Usage Confidence} & \textbf{Types of Apps} & \textbf{App Usage (hrs/week)}  \\  
        \hline \hline
        A1 & 74 & Male & Polish/ Australian & Glasses & Smartphone, PC & High &  
        \multirow{7}{3cm}{\\Video calls: 7/7 \\ Social media: 7/7 \\ Games: 6/7 \\ Banking: 5/7 \\ Shopping: 5/7 \\ E-books: 5/7 \\ Streaming: 5/7 \\ Audiobooks: 5/7 \\ Navigation: 4/5 \\ e-Health: 2/7} &  
        \multirow{7}{1.5cm}{4-9H: 3/7 \\ 10-20H: 1/7 \\ \textgreater 20H: 3/7} \\ 
        \cline{1-7} 
        A2 & 74 & Female & Australian & Glasses & Smartphone, Tablet, PC & Average &  &  \\  
        \cline{1-7}
        A3 & 78 & Female & Australian & Glasses & Smartphone, Tablet, PC & Average &  &  \\  
        \cline{1-7}
        A4 & 69 & Female & Maltese/ Australian & Glasses & Smartphone, Tablet, PC & High &  &  \\  
        \cline{1-7}
        A5 & 74 & Female & Australian & Glasses & Smartphone, Tablet, PC & Low &  &  \\ 
        \cline{1-7}
        A6 & 75 & Female & Australian & Glasses & Smartphone, Tablet, PC & Average &  &  \\  
        \cline{1-7}
        A7 & 67 & Female & Australian/ New Zealander & Glasses & Smartphone, Tablet, PC & Very High &  &  \\ 
        \hline \hline
        B1 & 72 & Female & Singaporean & Glasses, Hearing aids & Smartphone & Average & Shopping, Banking, Social media, Navigation, Streaming, E-books, Audiobooks & 2-3H \\
        \hline
        B2 & 74 & Male & English & Glasses, Hearing aids & Smartphone, Tablet & Low & E-books, Games & 4-9H \\
        \hline
        B3 & 65 & Female & Australian & Glasses & Smartphone, Tablet, PC & Very High & Shopping, Banking, Social media, Navigation, Video calls, Streaming & 10-20H \\
        \hline
        B4 & 82 & Female & Australian & Glasses, Hearing aids, Walking stick & Smartphone, Tablet & Low & Banking, Navigation & 2-3H \\
        \hline
        B5 & 80 & Female & English/ Australian & Glasses & Older mobile phone & Very Low & Phone calls & \textless 1H \\
        \hline
        B6 & 67 & Female & Australian & Glasses & Smartphone, Tablet & Average & Banking, Social media, Video calls, Navigation, Streaming, E-books, Audiobooks & 10-20H \\
        \hline
        B7 & 86 & Female & n/a & Glasses, Walking stick & Smartphone, Tablet & Very Low & n/a & \textless 1H \\
        \hline
        B8 & 71 & Male & Scottish & Glasses & Smartphone, Tablet, PC & High & Shopping, Banking, Social media, Video calls, Navigation, Streaming, E-books, Audiobooks, Learning & 4-9H \\
        \hline
        B9 & 75 & Male & English & Glasses, Hearing aids & Smartphone & Average & Social media, Navigation & 2-3H \\
        \hline 
    \end{tabular}
    }
    \caption{Exploratory focus group study participant demographic data. There were \(N_A = 7\) participants (A1--A7) in the pilot focus group study (FG-A), and \(N_B = 9\) senior participants (B1--B9) in the refined focus group study (FG-B). In FG-A, the questions regarding \textit{Types of Apps} used by seniors and \textit{App Usage} time per week were posed as \textit{raise-of-hands} questions during the session. Consequently, these data points could not be attributed to individual participants. In FG-B, these questions were integrated into the demographic data collection form, allowing for participant-specific data to be recorded.}
    \label{tab:exploratory_demographic_data}
\end{table}

\subsubsection{\textbf{Preliminary Persona Corpus}}

Based on our thematic analysis, we identified that seniors require three types of critical app adaptations to mitigate the accessibility barriers they face and better personalise their app usage experience: (1) presentation adaptations, (2) multi-modality adaptations, and (3) navigational adaptations. Thereafter, we iteratively derived three personas from our understanding of senior age-related accessibility pain points to contextualise these app adaptation categories. The first persona in our corpus visualised how vision-related limitations could hinder the user experience of seniors. The second persona focused on how mobility limitations, such as arthritis, could negatively impact the senior experience, coupled with other age-related limitations. Finally, we used the last persona to contextualise how cognitive issues, such as age-related memory limitations, could impact the senior UI experience and how seniors' attempts at personalisation through currently available mechanisms may have negative consequences.

These personas directly impacted how we designed our domain-specific languages and app adaptation features later in the design and development phase of our study. Furthermore, our evaluative user study with seniors in phase 3 further expanded our understanding of senior accessibility pain points, resulting in a refined set of personas that are presented throughout Section \ref{sec4:problems and solutions} of the paper.

\subsection{Design and Development of \toolname} \label{adaptforge}

\begin{figure}
\includegraphics[width=0.8\textwidth]{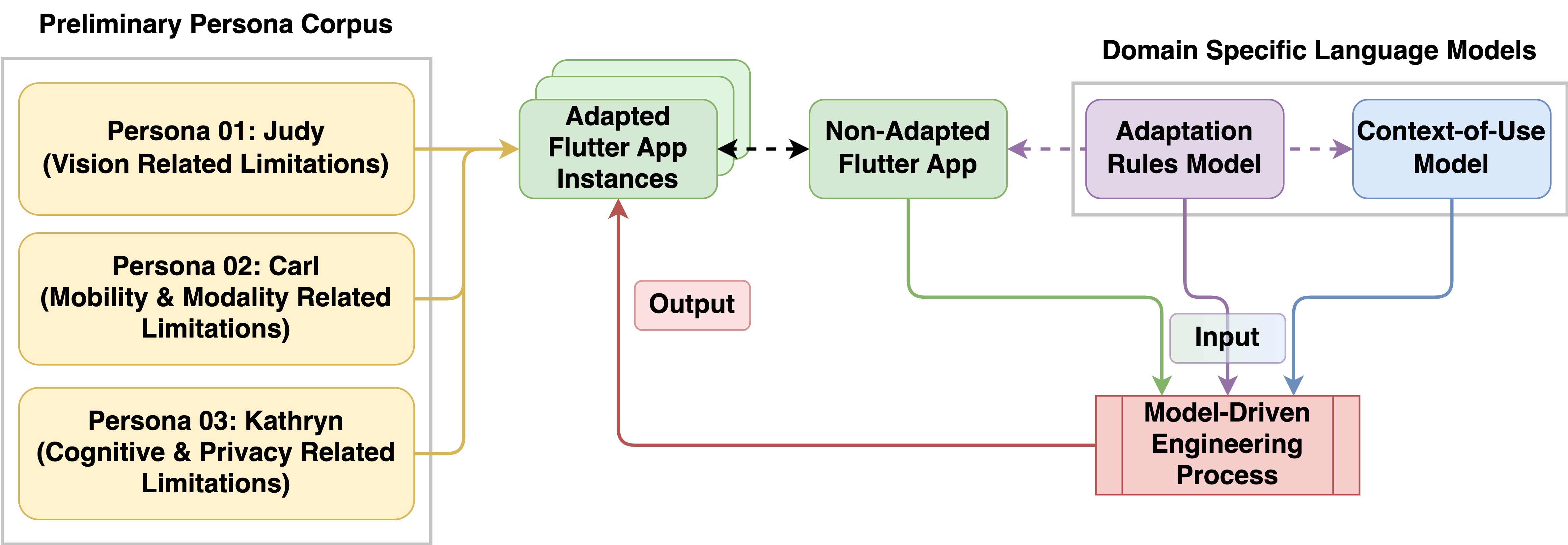}
\caption{An overview of the \toolname~tool that is being used to generate adapted app UI instances based on the preliminary persona corpus}\label{tool_overview}
\Description{This figure illustrates the abstract methodological framework of the AdaptForge tool. The model-driven engineering process begins with a non-adapted Flutter application and incorporates two domain-specific language models: the Adaptation Rules Model and the Context-of-Use Model. Based on the preliminary persona corpus identified during the exploratory phase, the tool generates adapted instances of the original Flutter application to address specific accessibility and personalisation needs.}
\end{figure}

Using the preliminary personas identified in the Exploratory phase, we developed our \toolname~ adaptation platform \cite{wickramathilaka2025}, as illustrated in Figure \ref{tool_overview}. \toolname~ contains three key components. First is ContextDSL, a domain-specific modelling language capable of capturing a diverse range of age-related parameters of a senior via a Context-of-Use model. By \textit{context-of-use}, we mean the n-tuple of contextual dimension categories: (1) \textbf{User context:} user-specific details such as impairments, preferences, technological skill/confidence level, and language comprehension level; (2) \textbf{Platform context:} variables related to hardware and software components the user interacts with, such as screen dimension, device type, operating system, and available input/output features; and (3) \textbf{Environment context:} parameters related to the environment in which the users interact with the system (i.e., ambient light, ambient noise, and geo-location) \cite{wickramathilaka2025, calvary2002, Stephanidis2000}.
The next component is our AdaptDSL, a domain-specific modelling language, which enables us to use the contextual parameters provided in the context-of-use model to express concrete adaptation operations into a Flutter app's source code through conditionally applied Adaptation Rules. These two DSLs can express the accessibility and adaptation needs of an individual senior or a group of seniors and adapt an app similar to the examples provided in Figures \ref{presentation ui}, \ref{navigation ui}, and \ref{modality ui}.
Finally, to ensure that the accessibility and adaptation needs modelled through our DSLs are applied to the app's source code, we developed a Model-Driven Engineering process pipeline that takes the non-adapted source code of a Flutter app along with our DSL models to output adapted app instances to suit the senior needs modelled.

\subsection{Evaluative Phase}

After implementing our \toolname~prototype and its adaptive app generation workflow, we re-engaged with the same U3A senior community that had initially provided insights into their accessibility needs. The objective was to present our approaches and obtain feedback on how our solution addressed their identified challenges. This phase also enabled us to showcase high-fidelity prototypes, facilitating more focused input from participants regarding their expectations for app personalisation in mitigating age-related limitations.

The evaluative focus groups and subsequent data analysis served both to assess our current understanding of senior accessibility requirements and to further extend this understanding. As a result, we refined the existing persona corpus into a more representative and nuanced set of personas, as illustrated throughout this paper. Finally, to address the identified accessibility issues, we leveraged both the empirical evidence obtained from the prototype evaluation and established knowledge from the UI accessibility literature and guidelines. This informed the development of targeted accessibility and personalisation recommendations for developers designing applications for senior users.
We also conducted an interview study with Flutter developers~\cite{flutter} to evaluate the usefulness, usability, and practical feasibility of the \toolname~tool. The findings of this developer tool evaluation are presented in detail in Wickramathilaka et al.~\cite{wickramathilaka2025}.

\subsubsection{\textbf{Evaluative Focus Groups: Design}}

We first obtained approval from the Monash Human Subject Ethics Committee (MUHREC Project ID: 42470). We then conducted a series of focus groups using a presentation-based approach. To facilitate discussions, we presented questions and demonstrated the application through short video clips\footnote{\href{https://drive.google.com/drive/folders/1_Eooe_XxzAHi8cfUHfWHP8WB4SyP55fs?usp=sharing}{Google drive link to video demos}}. Although we were capable of demonstrating adapted apps on a mobile device (in a debug environment), we opted to screen record different adapted versions of the same app\footnote{\href{https://github.com/adeeteya/FlutterFurnitureApp}{Open source app Github repository}} running on a real mobile device for demonstration convenience. These demonstrations illustrated a typical UI workflow, including user login, product browsing, and navigation to a form page for adding a shipping address.

The focus group session was structured around three key adaptation types: presentation, multi-modality, and navigation. Each adaptation was introduced through a video demonstration, beginning with the app in its original, non-adapted form, where participants provided feedback on usability and accessibility challenges. The second demonstration showcased presentation adaptations, where \toolname~was used to enhance text size, input box borders, and contrast, allowing participants to compare the adapted version with the original (Figure \ref{presentation ui}). The third demonstration introduced multi-modality adaptations, incorporating text-to-speech and speech-to-text features to improve interaction with the shipping address form  (Figure \ref{modality ui}). The final demonstration focused on navigation adaptations, transforming the multi-field form into a step-by-step wizard to simplify the user experience (Figure \ref{navigation ui}). The session concluded with a discussion on app run-time adaptations, gathering insights for future improvements.

\subsubsection{\textbf{Evaluative Focus Groups: Data Collection}}

We held three focus group sessions in person, each lasting approximately an hour. The sessions were conducted by the authors, audio recorded, and then transcribed using Otter.ai. During data collection, we inquired about any accessibility barriers or enablers that participants encountered while using the prototype UIs.

\subsubsection{\textbf{Evaluative Focus Groups: Data Analysis}}

First, we manually cleaned the transcripts generated from Otter.ai to ensure they accurately reflected participants' statements. Thereafter, we conducted an initial coding of these transcripts to generate the first iteration of codes. In this phase, we specifically examined accessibility barriers and enablers encountered by seniors, but in relation to the prototype used as a demonstrative tool. This approach differed from our earlier study, which explored the same concepts from a broader and more general perspective. The data and initial codes underwent multiple iterations until they formed a meaningful hierarchy of themes and sub-themes.

\subsubsection{\textbf{Evaluative Focus Groups: Participants}} \label{evaluative_participants}

\begin{table}
    \centering
    \renewcommand{\arraystretch}{1.3} 
    \scriptsize
    \resizebox{\textwidth}{!}{
    \begin{tabular}{|l|c|c|>{\raggedright\arraybackslash}p{1.3cm}|>{\raggedright\arraybackslash}p{1.5cm}|>{\raggedright\arraybackslash}p{1.5cm}|>{\raggedright\arraybackslash}p{2cm}|>{\raggedright\arraybackslash}p{2cm}|>{\raggedright\arraybackslash}p{4cm}|>{\raggedright\arraybackslash}p{1.5cm}|}
        \hline
        \textbf{\centering ID} & \textbf{Age} & \textbf{Gender} & \textbf{Culture} & \textbf{Occupation (prior)} & \textbf{Usage of aids} & \textbf{Mobile device usage} & \textbf{App usage confidence level} & \textbf{Types of Apps} & \textbf{App usage (hours-per-week)} \\ \hline \hline
        C1  & 74 & F & Australian & Public servant & Eye glasses & Android phone, Tablet & Average & Banking, Shopping, Social media, Navigation & 4-9 \\ \hline
        C2  & 85 & M & Anglo & Engineer & Eye glasses, Hearing aids & - & Very low  & - & \textless 1 \\ \hline
        C3  & 75 & M & Australian & - & Eye glasses & iPhone, Tablet & High  & Banking, Shopping, Social media, Navigation, Streaming, E-books, Audiobooks & 10-20 \\ \hline
        C4  & 72 & F & Australian & Nurse & Eye glasses & Android phone & Average  & Shopping, Social media, Navigation, E-books & 4-9 \\ \hline
        C5  & 76 & F & Indian & Educator, Academic & Eye glasses & iPhone, Tablet, Kindle & Average  & Banking, Shopping, Social media, Navigation, E-books & 2-3 \\ \hline \hline
        D1  & 68 & F & English & Academic librarian & Eyeglasses & iPhone, Tablet & Very high  & Banking, Shopping, Navigation, E-books, Audio-books & 4-9 \\ \hline
        D2  & 69 & F & Australian & Dental assistant & Eye glasses & Android phone & Average  & Banking, Shopping, Social media, Navigation, Streaming, E-books, Learning & \textgreater 20 \\ \hline
        D3  & 72 & F & Australian & Nurse & Eye glasses & iPhone, Tablet & Average & Banking, Shopping, Social media, Navigation, Streaming & 10-20 \\ \hline
        D4  & 83 & M & Australian & Business owner & Eye glasses, Hearing aids & Android phone & Average  & Voice calls & \textless 1 \\ \hline
        D5 & 66 & F & Australian & Marketing & Eye glasses & iPhone & High  & Banking, Shopping, Social media, Navigation, Streaming, E-books, Audio-books, Games & \textgreater 20 \\ \hline
        D6 & 63 & F & Dutch & Community support & Eye glasses & Android phone & Average  & Social media, Browsing & 10-20 \\ \hline
        D7 & 62 & F & Chinese & - & n/a & Android phone, Tablet & High  & Banking, Shopping, Social media, Navigation, Streaming, E-books, Audio-books & 2-3 \\ \hline
        D8 & 75 & F & Australian & Casual librarian & Eye glasses & iPhone, Tablet & High  & Banking, Shopping, Social media, Navigation, E-books & \textgreater 20 \\ \hline \hline
        E1 & 80 & M & German & Industrial sales & Eye glasses, Hearing aids & iPhone, Tablet & Average & Banking, Social media & \textless 1 \\ \hline
        E2 & 68 & F & Australian & Trainer, Assessor & Eye glasses & iPhone & Average to low & Banking, Shopping, Social media, Navigation, Streaming, E-books, Audio-books, Email & 10-20 \\ \hline
        E3 & 61 & F & Australian & Information management officer & Eye glasses & Android phone, Tablet & Average  & Banking, Navigation, Audio-books & 10-20 \\ \hline
        E4 & -  & F & Indian & Admin officer & Eye glasses & Android phone & Average & Banking, Social media, Streaming & 4-9 \\ \hline
        E5 & 70 & F & Australian & Banking & Eye glasses & Android phone, Tablet & Average  & Banking, Shopping, Social media, Navigation, E-books & 10-20 \\ \hline
        E6 & 68 & M & Australian, Irish & Public servant & Eye glasses & Android phone & High  & Navigation, Email, Voice calls, Messaging & 2-3 \\ \hline
        E7 & 79 & F & Australian & Funeral director & Eye glasses & Android phone, Tablet & Average  & Banking, Shopping, Social media, Navigation, Streaming, E-books & \textgreater 20 \\ \hline
        E8 & 76 & F & Australian & Service assistant & Eye glasses & iPhone, Tablet & Average & Shopping & \textgreater 20 \\ \hline \hline
        F1 & 73 & M & Scottish & Senior manager IT & Eye glasses & Android phone, Tablet & High  & Navigation, Messaging & 2-3 \\ \hline 
    \end{tabular}
    }
    \caption{Evaluative focus group study senior participant demographic data. A total of \(N_{CDEF} = 22\) seniors participated in this study stage, which comprised three focus group sessions. The first group (FG-C) included 5 participants (C1--C5), the second group (FG-D) had 8 participants (D1--D8), and the third group (FG-E) comprised 8 participants (E1--E8). Additionally, F1 attended all three sessions as an active participant and assisted in facilitating the discussions at the University of the Third Age (U3A).}
    \label{tab:evaluative_demographic_data}
\end{table}

We recruited 22 senior participants from a University of the Third Age (U3A) chapter in Australia. Each participant was gifted a 30 AUD voucher for a local supermarket as a token of appreciation.
The demographic information of the senior participants in our evaluative focus group study is summarized in Table~\ref{tab:evaluative_demographic_data}. The average age of the participants was 72.1 years, with a median age of 72 years. More than three-quarters of the participants were women (77.3\%, \(n=17\)). Culturally, almost all participants identified as Australian or European, with the former comprising 63.6\% (\(n=14\)) of the sample.  
There were no significant patterns in prior occupations, reflecting varied skills, experiences, and backgrounds among participants. Nearly all participants reported having some form of vision impairment, as 95.5\% (\(n=21\)) stated that they use spectacles. Additionally, all participants owned a smartphone, with the majority using Android devices (59.1\% (\(n=13\))). A similar proportion (59.1\% (\(n=13\))) also owned a tablet.  

When asked about their confidence in using mobile apps, participants rated themselves on a Likert scale ranging from very low to very high. Only 9.1\% (\(n=2\)) reported having below-average confidence. The majority (59.1\%, \(n=13\)) rated their skills as average.  
To assess app usage patterns, we filtered the most popular app domains identified in our exploratory study and incorporated them into the demographic data collection. The results aligned with our initial findings, revealing the most commonly used app categories among participants. Notably, navigation apps were the most frequently used, with 66.7\% (\(n=16\)) of participants relying on them. Online banking and social media applications were also widely adopted, each used by 62.5\% (\(n=15\)) of participants. Similarly, online shopping was prevalent among 58.3\% (\(n=14\)) of participants. Additionally, e-books were used by 45.8\% (\(n=11\)), while video streaming apps were utilised by 33.3\% (\(n=8\)). Audiobooks were the least commonly used, with only 20.8\% (\(n=5\)) of participants engaging with them. These findings underscore the diverse digital engagement of seniors and highlight key app categories that must prioritise accessibility enhancements.  
Only 8.3\% (\(n=2\)) of participants reported not using any of these app domains. Finally, 50\% (\(n=11\)) of the participants reported using mobile applications for 10 or more hours per week.

\subsubsection{\textbf{Refined Persona Corpus and Accessibility Recommendations}}

Scattered throughout this paper, we present a set of eight personas that reflect the real-world needs of our 38 senior participants across both the exploratory and evaluative phases of our study (Figures~\ref{persona example}, \ref{judy_persona}, \ref{kathryn_persona}, \ref{liam_persona}, \ref{usha_persona}, \ref{ava_persona}, \ref{stefan_persona}, \ref{dorothy_persona}). The participant statements that were used to derive these personas are detailed in Appendix Section~\ref{persona evidence}.
Each persona illustrates one or more themes discussed in Section~\ref{sec4:problems and solutions}. Within each theme, we begin by using a persona to contextualise a specific accessibility barrier experienced by seniors. We then follow this with a set of accessibility recommendations to address the highlighted barrier. These recommendations are grounded in insights from our senior focus group studies, established accessibility standards and guidelines, and platform-specific developer resources.
In Section~\ref{recommendations}, we provide a consolidated list of these recommendations, grouped by the same thematic structure used in Section~\ref{sec4:problems and solutions}, to offer readers a summarised and actionable reference.

\section{Problems and Solutions} \label{sec4:problems and solutions}

In this section, we synthesise the analysis conducted in both our exploratory and evaluative studies into a set of Accessibility Problem Dimensions (APD) faced by seniors. Additionally, we propose solutions from a software developer's perspective to address these issues. These solutions are derived from our findings in the evaluative focus group study and insights from prominent accessibility standards.

\subsection{APD 1: Text Readability} \label{text reability}

\begin{figure}
\includegraphics[width=\textwidth]{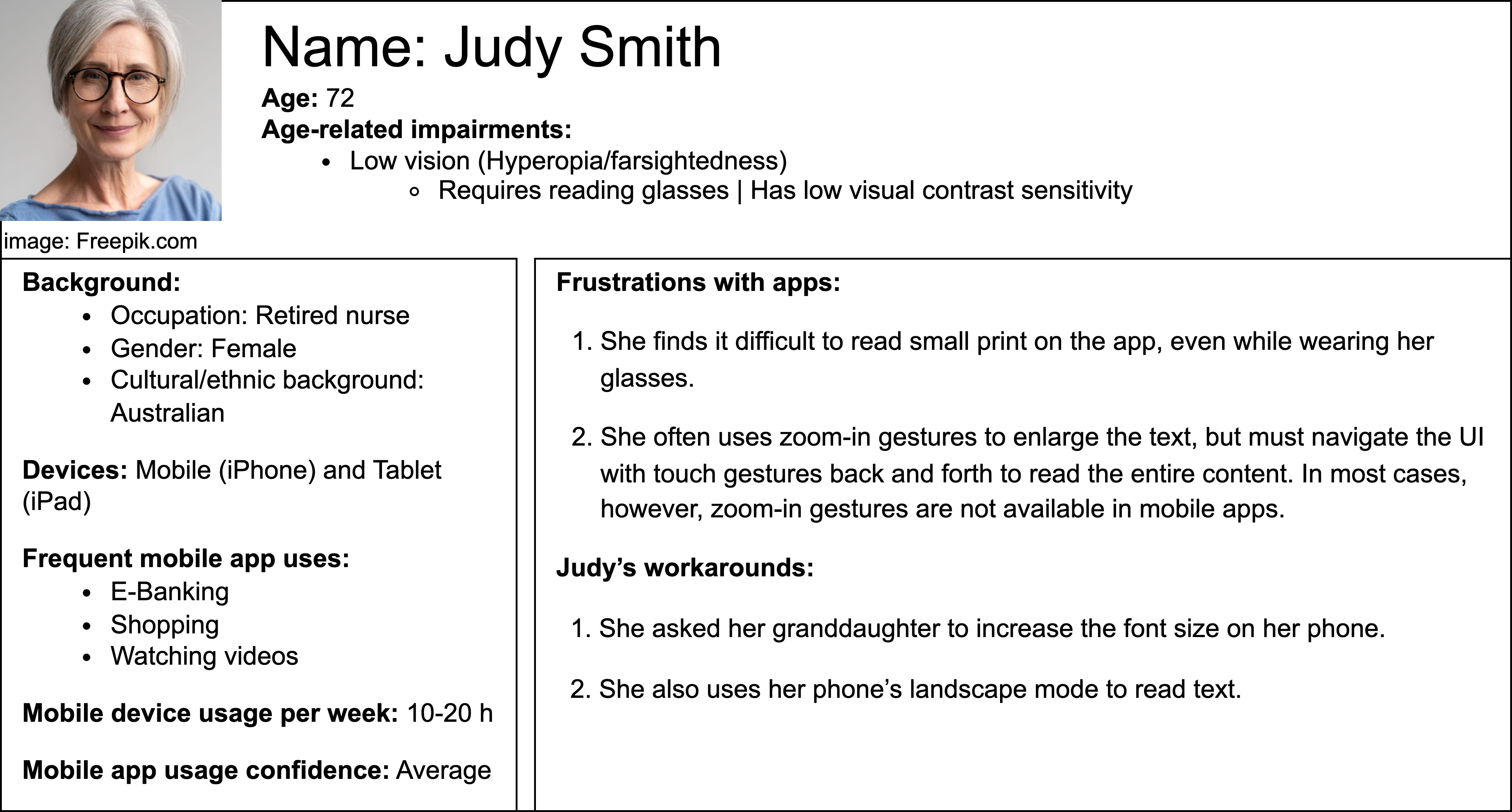}
\caption{A persona representing a fictional senior user, Judy, illustrates the challenges faced due to low vision, which creates accessibility barriers when attempting to read text on her applications.}\label{judy_persona}
\Description{This persona describes a fictional senior user, Judy Smith, aged 72. Judy has low vision and limited sensitivity to colour differentiation. Her confidence in using mobile applications is moderate. She experiences frustration when reading small text on apps. While she utilises zoom gestures on web applications to enlarge text, she finds that most mobile apps lack support for universal zoom functionality. As a workaround, she often relies on her granddaughter to adjust her device’s accessibility settings and frequently switches to landscape mode to improve text readability on her phone.}
\end{figure}

\subsubsection{\textbf{Problem:}}

Consider the persona given in Figure \ref{judy_persona}. It illustrates the accessibility barriers faced by a 72-year-old senior named Judy, who struggles to read text on her mobile apps due to her low vision condition. Across both phases of our qualitative study (FG-A, FG-B, FG-C, FG-D, and FG-E), we encountered similar frustrations and workarounds employed by seniors to mitigate these issues.  

For instance, like our fictitious Judy persona, senior participants expressed difficulties reading text on smaller mobile device screens. One participant from our evaluative focus groups (FG-D) articulated this frustration:  

\begin{quote}
    \textit{"When something is really small, I have to go open over and over to make it bigger, bigger, bigger so I can actually read."}
\end{quote}  

This highlights a common strategy discussed by many senior participants across the exploratory focus groups (FG-A and FG-B): employing pinch-out touch gestures to zoom in on text to improve readability. Several participants in these groups noted that while zoom-in gestures can be helpful, they also introduce additional challenges. When zoomed in, seniors can only view a portion of the full-text width, requiring them to continuously navigate the interface using vertical, diagonal, and horizontal touch gestures to read the content seamlessly.   

Yet another strategy that we encountered in three focus group sessions across both the exploratory and evaluative phases (FG-B, FG-C, and FG-E) was seniors using accessibility settings on their mobile devices to enhance text readability. However, it is worth noting that only a few participants adopted this option, while others appeared unaware of or unwilling to make such configuration changes. Based on our observations, this strategy was more prevalent among the more tech-savvy participants with high or very high levels of self-confidence in using computing devices and software. Additionally, seniors who had a close family member or friend willing to configure accessibility settings were more likely to benefit from these enhancements. 

\subsubsection{\textbf{Our Solution}}

\begin{figure}
\includegraphics[width=0.7\textwidth]{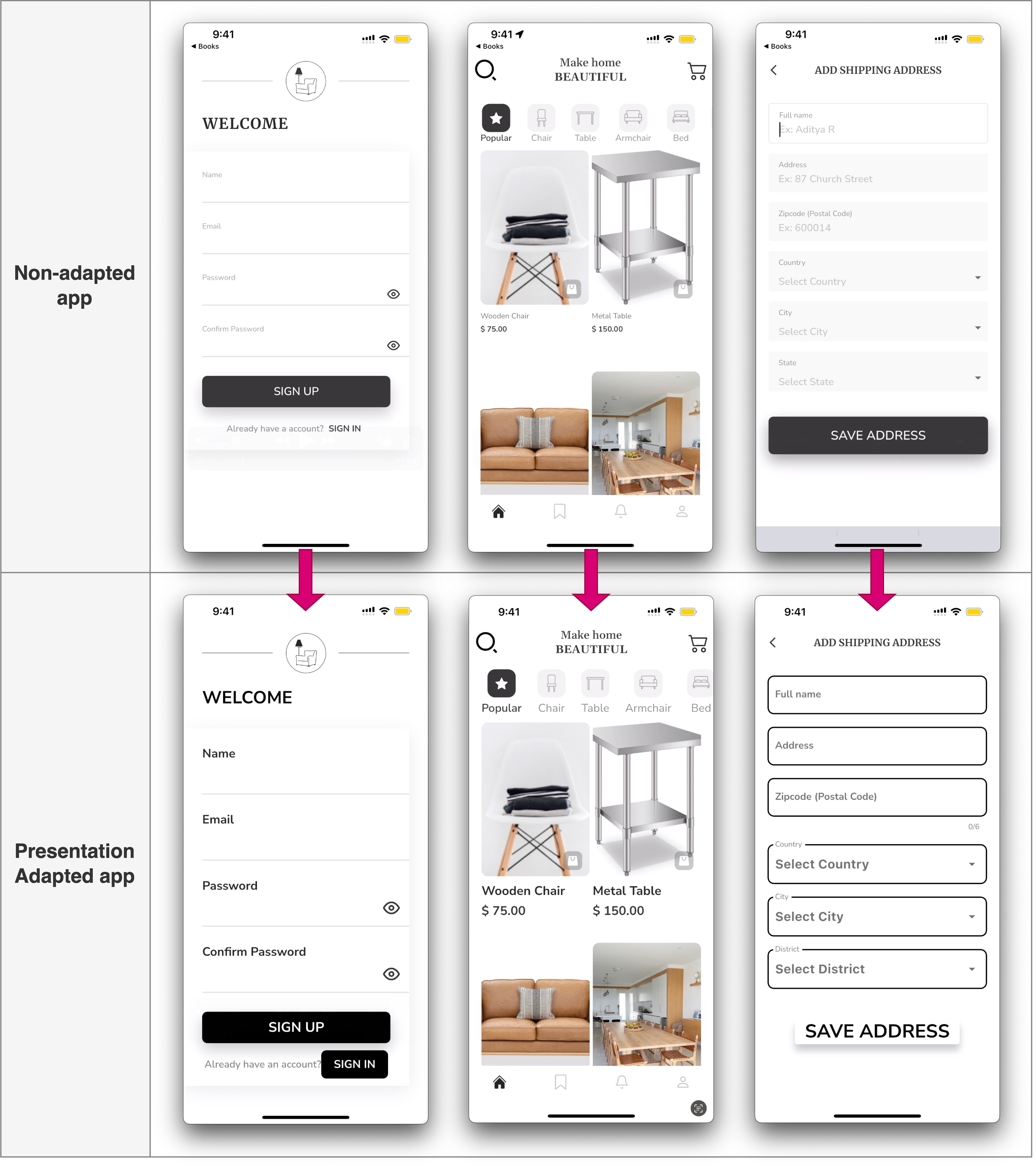}
\caption{This section presents adapted user interface examples derived from an initial set of non-adapted mobile app UIs. The adaptations include: (i) increasing body text size from 12pt to 24pt; (ii) applying a font weight of 700 to most body text; (iii) changing font colour from grey to black for enhanced readability; and (iv) enhancing the form fields in the \textit{Add Shipping Address} form with more prominent borders to improve visual distinction.}\label{presentation ui}
\Description{The figure presents a comparative view of three user interface examples from a non-adapted mobile application: a sign-up form, a home page, and an Add Shipping Address form. Each interface is shown in both its original and adapted forms. The adaptations applied include: (i) increasing body text size from 14pt to 24pt; (ii) applying a font weight of 700 to most body text; (iii) changing font colour from grey to black to improve readability; and (iv) enhancing the form fields in the Add Shipping Address form with more prominent borders to increase visual clarity.}
\end{figure}

During the evaluative phase of our study, we first demonstrated an unaltered instance of an open-source retail app (Figure \ref{presentation ui}) to seniors, with body text ranging from 12 to 16 pt. The majority of the body text had the 'normal' level of 400 text weight, as defined by the OpenType typography specification \cite{opentype}. When we presented this non-adapted app instance to seniors, all three evaluative groups (FG-C, FG-D, and FG-E) commented on its readability issues. For example in FG-E, the following was mentioned:  

\begin{quote}
    \textit{
    It looks a bit wishy-washy to me? [...] It's a bit insignificant. It's like... it doesn't stand out strongly. It's wishy-washy [text looks faint and weak].
    }
\end{quote}  

The term 'wishy-washy' effectively conveys our senior participants' perception of text readability in the non-adapted app instance. These participants found the generally used text sizes and weights for body text to be insufficient, often creating accessibility barriers. However, it was unanimously well received when we demonstrated the same app modified through our adaptation workflow, with adjustments such as a font size of 24 pt and a bold font weight of 700. 


\subsubsection{\textbf{Recommendations}}

Several studies~\cite{kurniawan2008, harte2018, morey2019, chai2017, watkins2014} advocate for the use of larger text in mobile applications designed for seniors. However, optimal font sizes for different text contexts, such as body and header text remain underexplored. The closest relevant guidance was found in Morey et al.~\cite{morey2019}, where they recommend a minimum font size of 30~pt for ‘critical text’ and 20~pt for ‘secondary’ text. Although the definitions of these categories are not explicitly stated, it can be inferred that ‘secondary’ text refers to non-header text that does not demand the user’s immediate attention. Our prototype aligns with this finding, as we used 24~pt for enhanced body text, which was positively received by participants.
Chai and Cao~\cite{chai2017} further observed that 27~pt (36~px) was optimal for seniors, based on character recognition speed. However, their study involved Chinese characters, which tend to be more complex than the English characters used in our prototypes.

Based on these findings and our empirical evidence, we \textbf{\textit{recommend a minimum font size of 20~pt for body text in mobile applications targeting senior users}}. Where possible, this minimum should also apply to smaller text elements such as footnotes and captions. Text elements requiring greater prominence, such as headers, titles, and critical information should scale upwards from this baseline. An example of such scaling is provided in Apple’s developer specifications for iOS and iPadOS, particularly in the \textit{xxLarge} dynamic type size category~\cite{iosDevGuidelines}. Additionally, \textbf{\textit{we recommend using a minimum font weight of 700}}, following the OpenType specification~\cite{opentype} on font weight, to further enhance readability for senior users.

An alternative yet complementary approach to enhancing text readability has been proposed by various studies and guidelines/standards \cite{Shamsujjoha2025, iso9241, androidDevGuidelines, iosDevGuidelines}. Rather than prescribing fixed font sizes, these resources advocate enabling users to adjust text size according to their preferences. For instance, the Web Content Accessibility Guidelines (WCAG) \cite{wcag2.2} recommend that text should be resizable up to 200\% without loss of content or functionality (WCAG Success Criterion 1.4.4). To contextualise this, consider how Carl (Figure~\ref{persona example}) and Judy (Figure~\ref{judy_persona}) utilise either global accessibility settings on their devices or app-specific settings to increase text size. In accommodating such configurations, \textbf{\textit{developers must ensure that user interface (UI) elements remain properly aligned and are not truncated or displaced off-screen}} \cite{iso9241}.

We \textbf{\textit{recommend that developers utilise specific font types when designing applications for older adults}}, namely serif or sans serif fonts such as Helvetica, Arial, and Times New Roman. These fonts are selected from the ReDEAP accessibility design pattern set \cite{redeap, ahmad2020}, which is grounded in existing studies and empirical research. It is important to note, however, that platform providers often specify their own preferred system font families. For instance, iOS/Apple recommends the use of the San Francisco (SF) sans serif and New York serif typeface families, advising developers to prioritise these over custom fonts due to their inherent support for dynamic, system-level accessibility configurations. In contrast, the use of custom typefaces may necessitate additional development and testing efforts to replicate similar adaptive text behaviours \cite{iosDevGuidelines}. The WCAG guideline 1.4.12 on text spacing can also be used to enhance text readability: It \textbf{\textit{recommends that line spacing be at least 1.5 times the font size, spacing between paragraphs be at least 2 times the font size, letter spacing be at least 0.12 times the font size, and word spacing be at least 0.16 times the font size.}}

Although developers rarely implement universal zoom gestures in mobile apps, if zoom functionality is necessary to enhance text readability, \textbf{\textit{we recommend that a magnifying glass widget/overlay should be considered}} \cite{iso9241}. These widgets allow users to zoom in on text without requiring seniors to navigate back and forth to read content.

\subsection{APD 2.1: Low Contrast and Use of Colour}

\begin{figure}
\includegraphics[width=\textwidth]{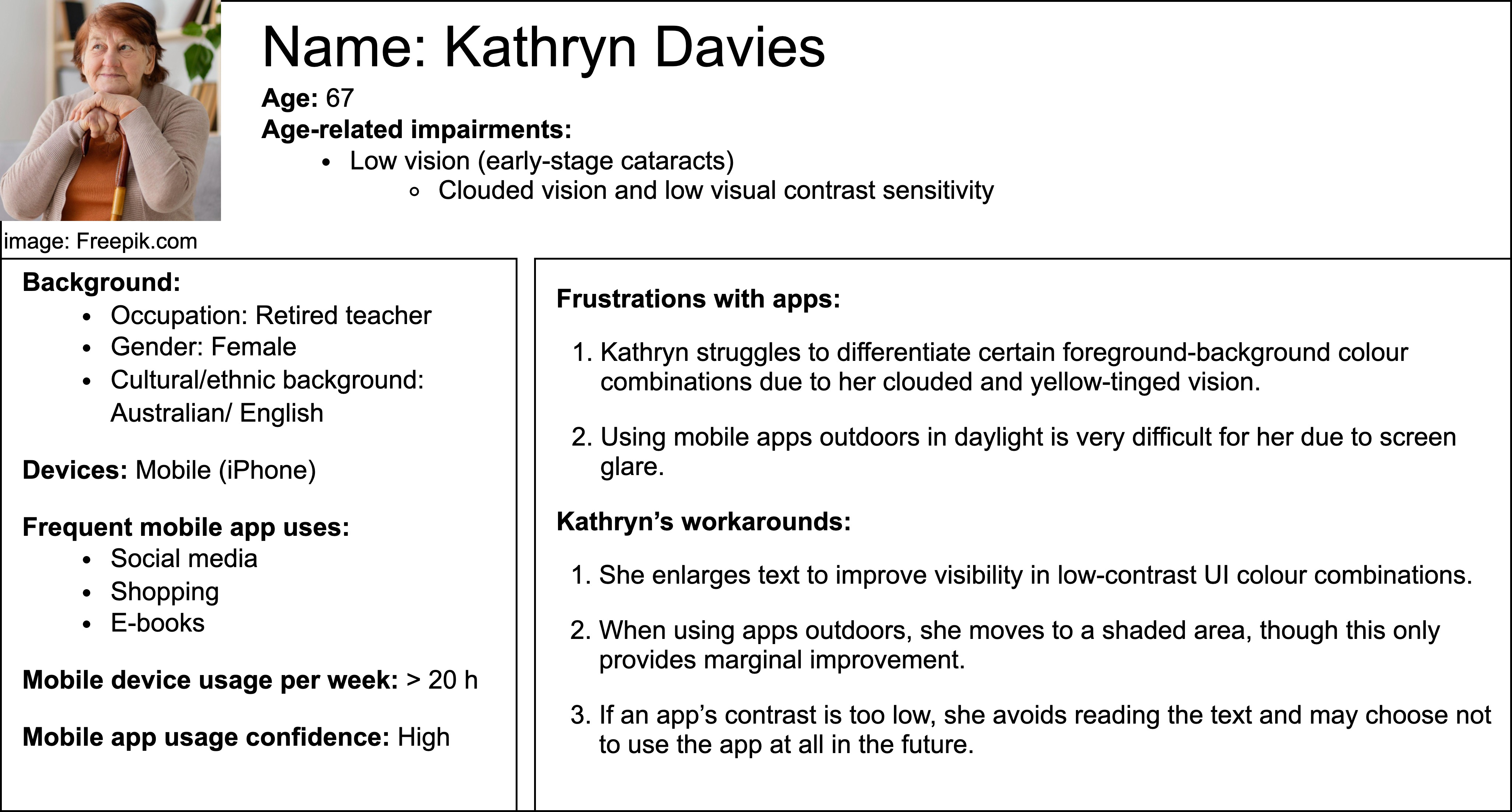}
\caption{A persona representing a fictional senior user, Kathryn, illustrates the challenges faced by seniors when encountering low contrast background-foreground colour combinations in UI.}\label{kathryn_persona}
\Description{The figure presents the fictional persona of Kathryn Davies, aged 67. Kathryn has low vision caused by cataracts, which result in a yellowish visual tint. Despite her condition, she is a frequent and confident mobile app user. She experiences significant difficulty distinguishing between foreground and background elements when colour contrast is low. Additionally, she struggles to read content outdoors due to screen glare. To address these challenges, Kathryn enlarges text to improve readability. When outdoors, she seeks shaded areas to mitigate glare. However, when app interfaces feature poor colour contrast, she often chooses to avoid using those applications altogether.}
\end{figure}

\subsubsection{\textbf{Problem:}}

Yet another key accessibility barrier found in all our focus group sessions (FG-A, FG-B, FG-C, FG-D, and FG-E) was insufficient contrast between background and foreground elements in app UIs. To illustrate this issue, we created a persona that encapsulates the pain points discussed by our participants (Figure \ref{kathryn_persona}).

Kathryn, a retired nurse, is confident in using mobile devices due to her frequent usage over the past few years. However, her vision has gradually deteriorated with age, and she was recently diagnosed with early-stage cataracts. While she currently manages this condition with spectacles, she will require surgery in the future. Her clouded and yellow-tinged vision makes it difficult to use apps that lack high contrast between foreground and background colours. 



When encountering UIs with insufficient contrast, Kathryn attempts to mitigate the issue by increasing text size using her device’s native accessibility settings. However, this is not always an effective solution when the inherent contrast between text and background remains too low. Our study participants from FG-B expressed similar sentiments:

\begin{quote}
\textit{"It [inadequate contrast] would put you off reading it [text]. Suppose I can enlarge it and spread it so you can see it more closely, but I don't know how I would change the colour of it."}
\end{quote}

Furthermore, the issue of low-contrast UIs is particularly pronounced in high ambient light conditions, such as outdoor daylight use, where screen glare further exacerbates readability difficulties. In such environments, Kathryn often finds herself squinting and struggling for extended periods to comprehend on-screen content. In response, she adopts practical workarounds, such as moving to a shaded area or using her hand as a temporary shield to reduce glare. However, these solutions provide only marginal improvement, as she still has to contend with low-contrast UI elements under less-than-ideal lighting conditions.

When asked whether they attempted to adjust their device settings, such as enabling dark or high contrast modes to address these issues, many seniors highlighted a significant challenge: adjusting these settings requires clear visibility of the UI, which is precisely what is compromised under screen glare conditions. 


\subsubsection{\textbf{Our Solution}}

One of the themes that we frequently encountered during our analysis was the preference of seniors to have a UI that is optimised for high contrast. In fact, it was mentioned in both of our exploratory focus group sessions (FG-A and FG-B) that the seniors prefer a simple black and white colour theme in place of more flashy and colourful themes. 


This later influenced when we were designing our app adaptation workflow, which resulted in the black and white app instance that we showed to participants (Figure \ref{presentation ui}). In addition to the theme change, we also made the borders around UI elements such as buttons and form fields more prominent. This was a well-received change amongst all evaluative stage participants (FG-C, FG-D, FG-E), especially compared to the feedback received for the default non-adapted app instance. For example, here is a statement from a senior participant during FG-C:

\begin{quote}
\textit{"When it is just black and white, I find that easier. And I think [for] people who are older, again, 90 and so on, the simpler, the better. The clearer the message, the better."}
\end{quote}

\subsubsection{\textbf{Recommendations}}

The need for high-contrast user interfaces (UIs) has been a recurring theme in numerous studies and design guidelines \cite{salman2018, androidDevGuidelines, wcag2.2, iosDevGuidelines, iso9241, Shamsujjoha2025, harte2017, watkins2014, jonsson2016, panagopoulos2015}. We \textbf{\textit{recommend that developers implement high-contrast, monochrome (e.g., black-and-white) themes}} that can be enabled by senior users as needed, based on feedback from our cohort of older adult participants.
Implementing a fully monochrome black-and-white theme may not always be feasible due to branding and app identity considerations. In such cases, we \textbf{\textit{recommend that developers ensure their chosen colour schemes comply with WCAG Success Criterion 1.4.6 on enhanced contrast}} \cite{wcag2.2, iosDevGuidelines}. This criterion mandates \textbf{\textit{a minimum foreground-to-background contrast ratio of 7:1 for body text}} and other small text elements. \textbf{\textit{For larger text, such as headings, a reduced ratio of 4.5:1}} is acceptable.

Beyond predefined high-contrast themes, we \textbf{\textit{recommend that developers provide older users with the ability to select a preferred theme or to create a custom one}} \cite{redeap, Shamsujjoha2025, iso9241}. When offering such theme selections,  \textbf{\textit{developers should introduce profiles tailored to common visual impairments, such as colour blindness, macular degeneration, cataracts, and low visual acuity.}} However, implementing customisable themes introduces complexity, particularly in preventing users from selecting colour combinations that do not meet WCAG 1.4.6 requirements. Failure to enforce these constraints could diminish users’ confidence in personalising their apps, an issue further discussed in Section~\ref{fear of mistakes}. For users requiring perssonlised themes, we \textbf{\textit{recommend that developers consider providing colour pickers that restrict choices to combinations or gradients meeting the minimum contrast ratio.}}

To support robust theme personalisation at the app level, we  \textbf{\textit{recommend that developers adopt a theme propagation and management mechanism}}, such as the \textit{Theme} widget in the Flutter framework \cite{flutter}.

Consistent with prior empirical research \cite{redeap, Shamsujjoha2025, morey2019},  \textbf{\textit{we recommend use of app-level colour theme personalisation over exclusive reliance on platform-level accessibility settings}}. App-level personalisation can improve familiarity and increase usage among senior users \cite{redeap}. However, Apple's developer guidelines caution against this practice due to potential conflicts with system-wide appearance settings \cite{iosDevGuidelines}. Despite this challenge, we argue that enabling the coexistence of app-specific themes and device-specific settings (e.g., dark and light modes) is a worthwhile effort.

Finally, it is critical to evaluate the UI’s visual experience with senior users under both app-defined and device-defined themes. \textbf{\textit{Testing should also account for various ambient lighting conditions}}, such as outdoor daylight and dim environments. Additionally, \textbf{\textit{testing across different mobile devices is necessary}}, as features like the True Tone display available in several iOS and iPadOS devices adjust the white point based on ambient lighting \cite{iosDevGuidelines}, potentially affecting contrast and resulting in inconsistent user experiences.

\subsection{APD 2.2: White Background} \label{white background}

\subsubsection{\textbf{Problem and Our Solution}}

A fortunate but unintentional accessibility enabler that we identified across two of our focus groups in the evaluative stage (FG-C and FG-D) was the white background of the open-source app used for demonstration (Figure~\ref{presentation ui}). Participants found the UI of the prototype clean and simple, partly due to its overall minimalistic design, lacking flashy or multi-coloured elements, and partly due to its white background. For example, one participant noted in FG-C:  

\begin{quote}
\textit{"I find white backgrounds, just looking generally on the internet, is clearer to see than when there's multi colour."}
\end{quote}  

\subsubsection{\textbf{Recommendations}}

Although the use of a white background was specifically discussed in two focus group sessions,  \textbf{\textit{we only tentatively recommend that developers adopt it as part of an application's overall theme}}. This recommendation is made with caution due to the lack of corroborating evidence in the existing literature and accessibility guidelines. While prior discussions have highlighted the importance of high-contrast user interfaces, the literature generally supports monochrome themes, such as black-and-white configurations, without explicitly favouring white backgrounds with black text over the inverse for seniors.

Nevertheless, participant feedback suggested a preference among older users for user interfaces where a lighter colour (e.g., white) is used as the background. Based on this, we suggest that developers prioritise lighter background colours (such as white) wherever feasible. However, we also stress the importance of including traditional ‘dark mode’ options, where darker background colours (e.g., black) are used. Providing both options within appearance customisations enables senior users to select the theme that best suits their individual visual comfort and preferences.

\subsection{APD 3: Touch Interaction Targets Being Too Small}

\subsubsection{\textbf{Problem}}

When a senior such as Carl (Figure \ref{persona example}) or Liam (Figure \ref{liam_persona}) attempts to perform touch-based UI interactions, it becomes a difficult task due to their Arthritis condition. During our app demonstrations, one group of participants (FG-C) pointed out that the buttons and input fields in our app were not large enough to accommodate a senior affected by a mobility impairment, such as Arthritis:

\begin{quote}
\textit{"I'm just thinking of an older person that may have either a rigidity: some types of arthritis [can cause the hands to be] very rigid; or the opposite, quite shaky [hands]. That can cause difficulties for people clicking into even those very clear boxes. It might be something [a problem] that arises for people with those issues."}
\end{quote}

\subsubsection{\textbf{Recommendations}}

In our prototype demonstrations, we improved the visibility and size of certain buttons. However, feedback from participants during the prototype evaluation phase indicated that the size and spacing of buttons and other touch targets require further consideration to enhance accessibility for senior users. 
Numerous guidelines and studies concur that a touch target size of \(44 \times 44\) dp is optimal for pointer interactions (including touch actions on mobile devices) for older adults \cite{wcag2.2, iosDevGuidelines, Shamsujjoha2025}. This recommendation appears to be based on WCAG Success Criterion 2.5.5: Target Size (Enhanced) \cite{wcag2.2}. However, other sources propose different minimum dimensions. For instance, the Android Developer Accessibility Guidelines suggest a minimum of \(48 \times 48\) dp for touch targets \cite{androidDevGuidelines}. Harte et al. \cite{harte2017} further recommend that button areas be no smaller than 200~mm\(^2\), equating to approximately \(89 \times 89\) dp for a square touch target.
While Harte et al.'s recommendation specifically targets older users, we find the WCAG, Android, and Apple guidelines to offer a more practical balance. Accordingly, we \textbf{\textit{recommend a minimum touch target size of 7--10~mm in both height and width}}, with a preference for larger targets whenever layout constraints permit \cite{androidDevGuidelines, Shamsujjoha2025}. For example, Apple advocates (in iOS) for full-width buttons wherever possible to improve accessibility \cite{iosDevGuidelines}.

Another important consideration is the spacing between touch targets. The Android accessibility guidelines \textbf{\textit{recommend a minimum spacing of 8~dp between adjacent interactive elements}} \cite{androidDevGuidelines}, which can help reduce accidental touches and enhance usability for older adults.

\subsection{APD 4: Limited Screen Size} \label{screen size}

\begin{figure}
\includegraphics[width=\textwidth]{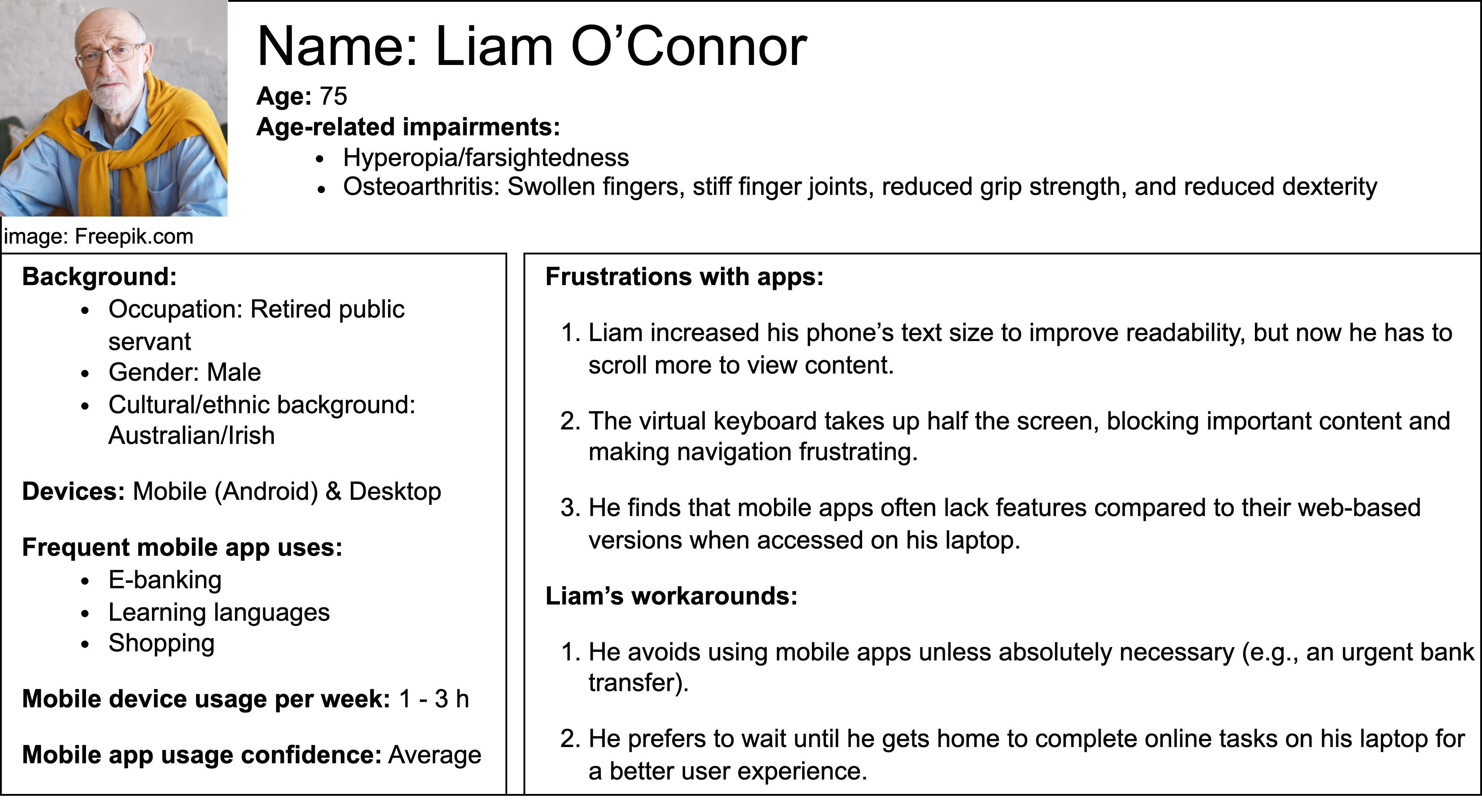}
\caption{A persona representing a fictional senior user, Liam, illustrates the challenges seniors face due to the limited screen size of mobile devices and the constraints of the apps running on them.}\label{liam_persona}
\Description{This persona describes Liam, a fictional senior user aged 75. Liam experiences far-sightedness, which impacts his ability to read text at close range. Additionally, he has osteoarthritis, resulting in swollen fingers, stiff joints, and reduced grip strength.}
\end{figure}

\subsubsection{\textbf{Problem:}}

Consider the persona of Liam, illustrated in Figure \ref{liam_persona}. Liam is a 72-year-old former public servant with farsightedness, requiring reading glasses. A more critical issue is his osteoarthritis, which limits his hand and finger dexterity as well as his grip strength. As a result, it is difficult for him to hold his phone for extended periods or perform fine touch gestures with his fingertips.  
Liam has increased the text size on his phone to improve readability; however, as a consequence, he now has to perform more frequent vertical scroll actions to view the full content of apps. Additionally, any input actions require the use of a virtual keyboard, which occupies nearly half of the screen space. This issue was raised in both focus groups during our exploratory study (FG-A and FG-B). For example, one participant explained:  

\begin{quote}
\textit{"My [partner] complains that when that [virtual keyboard] comes up, it can obliterate [partner's user experience]. Because it fills half the screen. It certainly causes [them] a lot of frustration. Then what [they] wanted to look at, [they] can’t see because the keyboard is there."}
\end{quote}

Even without considering the virtual keyboard, increasing text size results in less information being displayed on the screen at a given time. This further exacerbates the limitations of mobile apps, which often have a more restricted feature set compared to their desktop or web-based counterparts. One participant from FG-B highlighted this issue with their superannuation app, noting that its mobile version lacked features available in the web version. As a result, some participants from both focus groups mentioned that they would prefer to use their desktops or laptops for tasks rather than relying on the truncated mobile app experience. 

\subsubsection{\textbf{Our Solution}}

\begin{figure}
\includegraphics[width=0.6\textwidth]{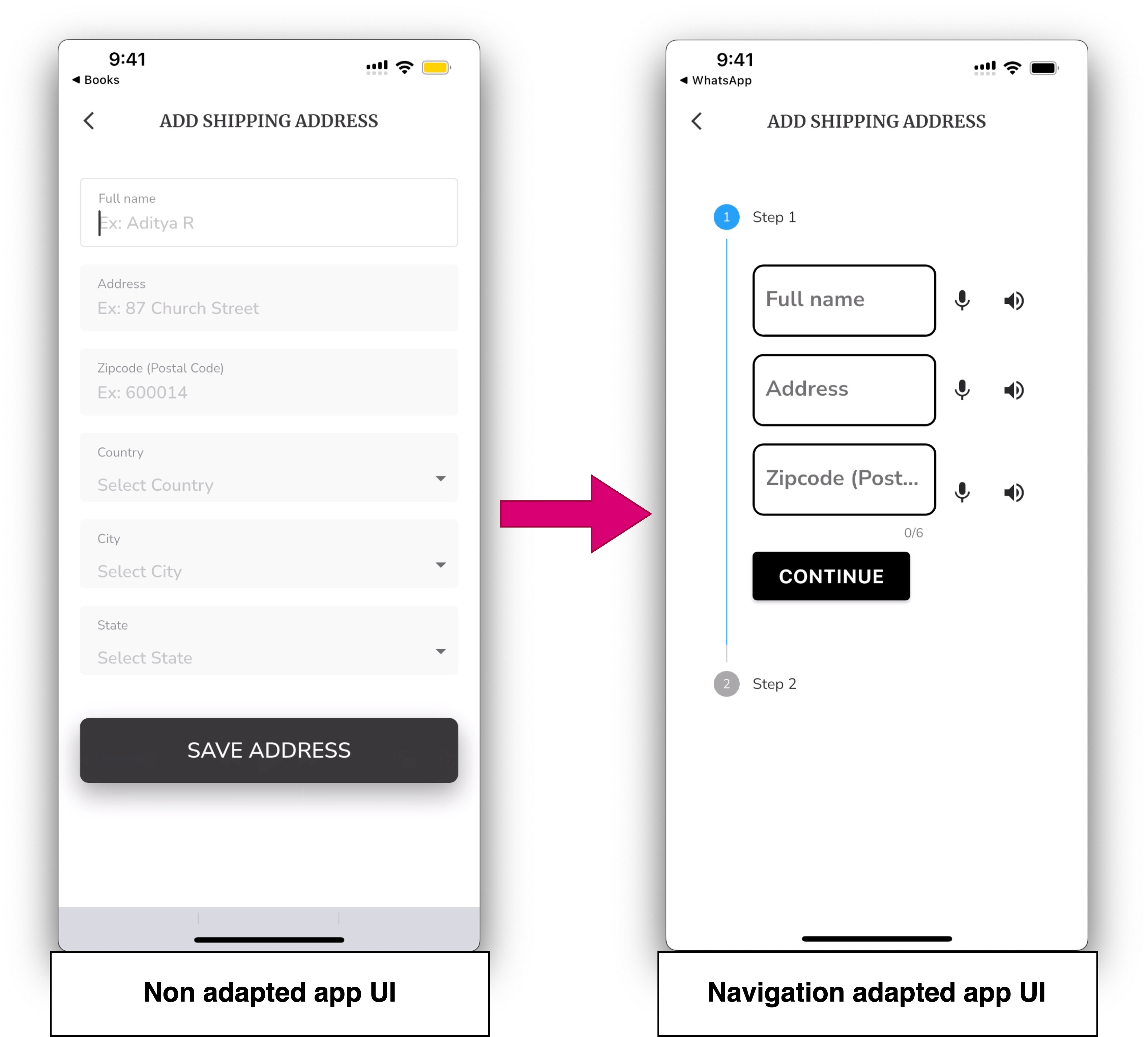}
\caption{Navigation adapted UI example. In this example, the non-adapted \textit{Add Shipping Address} form has been redesigned as a step-by-step interface, dividing the original form into two sequential steps.}\label{navigation ui}
\Description{In this example, the non-adapted Add Shipping Address form with has been redesigned as a step-by-step interface, dividing the original form into two sequential steps.}
\end{figure}

In our adapted app demonstration, we aimed to address general issues related to limited screen size and the additional content display challenges that seniors may face when increasing text size. This adaptation is illustrated in Figure~\ref{navigation ui}. Initially, the UI form contained multiple input fields and dropdown menus, presented as a single-page form requiring scrolling.  
When we increased text and form field sizes to enhance visibility, an unintended consequence emerged: seniors had to scroll even more than in the default app instance. To resolve this, we transformed the single-page form into a multi-page, wizard-patterned form, enabling users like Liam to progress through the form without excessive scrolling. Participants across all three focus groups in our evaluative phase (FG-C, FG-D, FG-E) expressed satisfaction with this adaptation. 

\begin{quote}
\textit{"I think, the fact that there is not a lot on the one screen that you got to go from screen to screen... that means that you don't feel overwhelmed with all of this stuff that you've got to fill out. So I think that is definitely an improvement."}
\end{quote}

\subsubsection{\textbf{Recommendations}}

The trade-off between increased UI element size (e.g., text and buttons) and the limited screen real estate available on mobile devices has been widely recognised in existing literature and platform guidelines \cite{iosDevGuidelines, androidDevGuidelines, harte2017, leitao2012, inostroza2016}. To address this issue, both iOS and Android development guidelines recommend designing responsive UI layouts that adapt to user actions such as switching between portrait and landscape orientations, adjusting font sizes, and accommodating content localisation (e.g., language changes) through rigorous testing \cite{iosDevGuidelines, androidDevGuidelines}. Nonetheless, these adaptive measures alone do not fully mitigate the drawbacks arising from personalisation actions taken by senior users, particularly as such changes often necessitate increased vertical scrolling to access content located at the bottom of the page.
To better accommodate senior users,  \textbf{\textit{we recommend an alternative approach whereby larger content sections are divided into logically structured groups}}. This ensures that, when users enlarge text or UI elements for improved readability or interaction, the content remains accessible. Such content can be presented in a sequential, \textbf{\textit{step-by-step wizard pattern}}, similar to the design shown in Figure~\ref{navigation ui}. This approach is supported by prior work, including the Android accessibility guidelines \cite{androidDevGuidelines}, Ahmad et al. \cite{redeap}, and Leitão and Silva \cite{leitao2012}, which advocate for layout reflow techniques that reduce the physical and cognitive burden on older users, particularly those with reduced dexterity or hand mobility.

An additional benefit of the wizard-based approach is its ability to minimise cognitive overload. When complex interfaces, such as those containing multiple widgets (e.g., text, images, input fields) are presented all at once, they may overwhelm older users. This concern was substantiated in our study and echoed in others \cite{redeap, inostroza2016}. For example, presenting a lengthy registration form in its entirety may be perceived as cognitively taxing by senior users. However, \textbf{\textit{if the form is broken down into manageable steps, such as ‘Step 1 of 3' to ‘Step 3 of 3', it can significantly enhance usability and reduce user frustration}} and even embed a sense of logical progression into a complex UI.

We also \textbf{\textit{recommend minimising the use of virtual keyboards wherever feasible.}} This suggestion aligns with findings from other studies \cite{redeap, barros2014}. As discussed further in relation to audio input and output modalities, virtual keyboards can occupy a substantial portion of the screen -- often nearly half -- thus exacerbating existing screen space limitations and reducing the overall usability of the interface for senior users.

\subsection{APD 5: Error Handling and Jargon} \label{errors}

\begin{figure}
\includegraphics[width=\textwidth]{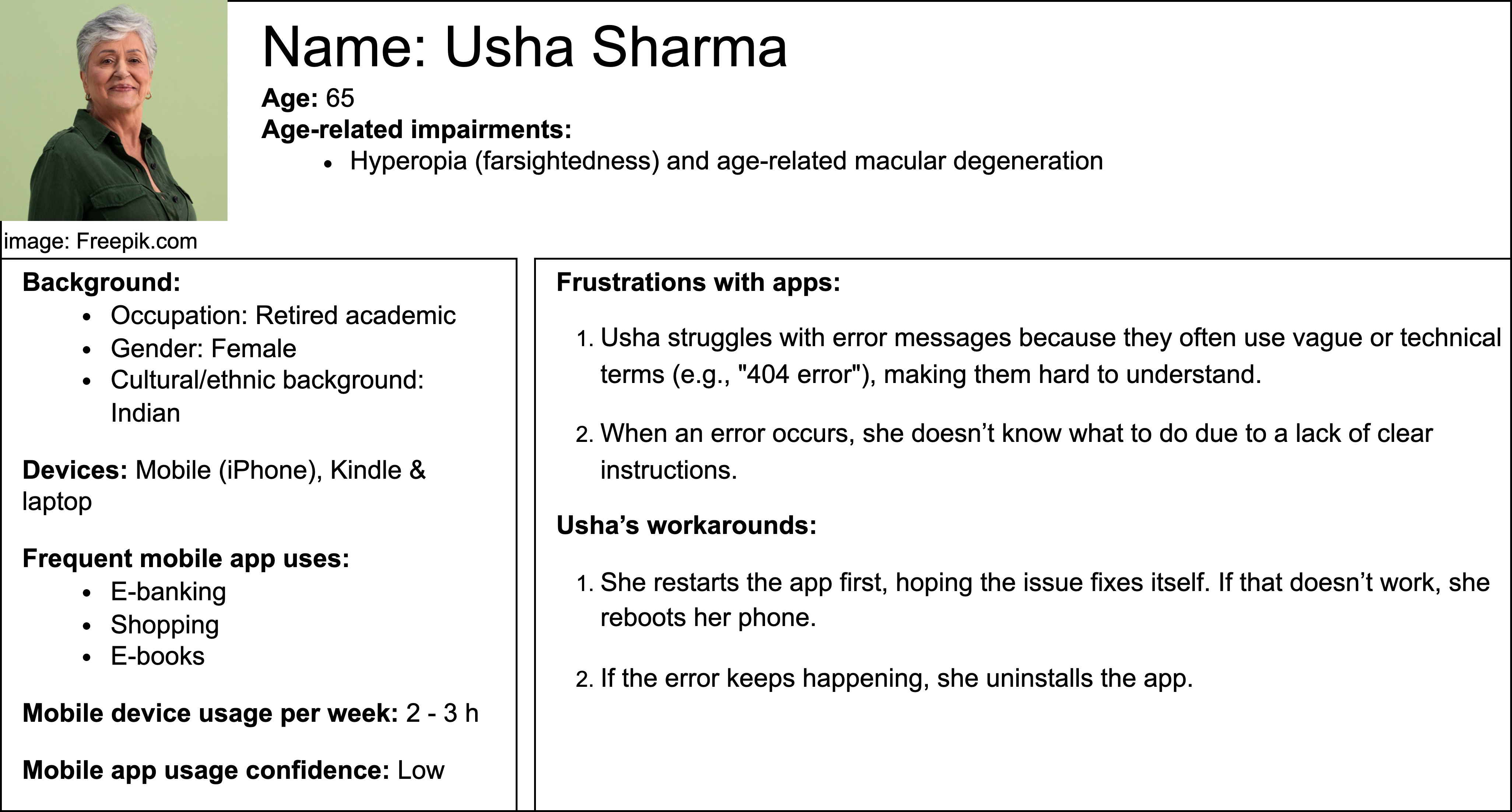}
\caption{A persona representing a fictional senior user, Usha, represents the issues faced by seniors when encountering vague and jargon-heavy error messages in their apps}\label{usha_persona}
\Description{The figure presents the fictional persona of Usha, a 65-year-old senior who demonstrates low confidence in using mobile applications.}
\end{figure}

\subsubsection{\textbf{Problem:}}

Consider the persona presented in Figure \ref{usha_persona}. Usha, a retired academic, recently encountered an issue while attempting to load an e-book into her preferred reading app. The application displayed an error message that she did not understand, as it lacked unambiguous language suited for users with limited experience and confidence in modern computing devices and applications.

Usha’s frustration reflects a recurring concern raised in both focus groups during the exploratory phase (FG-A and FG-B): seniors’ dissatisfaction with how applications communicate errors. A commonly mentioned strategy among participants was restarting the app or device, particularly among less tech-savvy seniors. In worst-case scenarios, unclear error messages may lead seniors like Usha to abandon or uninstall the app altogether 

Participants who demonstrated greater confidence in their computing skills expressed a willingness to seek solutions rather than abandon the application upon encountering an error message. Such individuals were observed in both FG-A and FG-B, in contrast to the majority who preferred to avoid engaging with app errors altogether. Nonetheless, it is essential that error messages present clear and intuitive resolution steps, avoiding technical jargon, dense text, and unfamiliar terminology that may overwhelm senior users. 

Another issue related to error messages and the overall UI of an app is the use of jargon, which often frustrates seniors. For example, one participant from FG-A expressed the following concern:  

\begin{quote}
\textit{"One of the most annoying things when you get an error message is they don't actually give you the message. They'll say error 404, what is error 404? So you've gotta go to the internet and type in error 404, and hope the error that they've given you is the same as the one you found on the internet. Why can't they actually tell you what the error is?"}
\end{quote}  

\subsubsection{\textbf{Recommendations:}}

The cognitive challenges associated with how errors are conveyed to senior users often through unfamiliar, vague terminology or technical jargon, is a complex user interface (UI) issue. Ineffective communication of critical error information has been highlighted in numerous studies and guidelines \cite{wcag2.2, iso9241, iosDevGuidelines, androidDevGuidelines, redeap, inostroza2016, salman2018, Shamsujjoha2025, harte2017, watkins2014, barbosa2015, almeida2015, barros2014, morey2019, jonsson2016}. 
To handle errors more gracefully, prior work recommends prioritising error prevention \cite{wcag2.2, redeap, inostroza2016}. This preventive approach is particularly relevant, as many senior users in our study demonstrated a tendency to ignore errors in the hope that the application would self-correct. While existing guidelines offer limited concrete strategies for implementing such a paradigm, we argue that it can be integrated into the application’s design architecture and \textbf{\textit{rigorously supported through quality assurance and bug-fixing processes during the software development lifecycle}}.

Our more realistic and immediately actionable recommendation is \textbf{\textit{to always communicate error messages in neutral, unambiguous language using layperson terms.}} Developers should avoid technical, cultural, and generational jargon, ensuring that the content of error messages is clear, direct, and easily comprehensible to older users \cite{wcag2.2, iso9241, iosDevGuidelines, androidDevGuidelines, redeap, inostroza2016, salman2018, Shamsujjoha2025, harte2017, watkins2014, barros2014}. Furthermore, \textbf{\textit{many senior users value step-by-step instructions that enable them to resolve issues independently}} \cite{redeap, salman2018, inostroza2016, Shamsujjoha2025}. Where appropriate, \textbf{\textit{help documentation and/or access to support personnel should be made easily accessible, with all guidance presented in jargon-free language}} \cite{barbosa2015, Shamsujjoha2025, inostroza2016, redeap, iso9241}.

These principles extend beyond error messaging to encompass all UI text. Applications intended for senior users should minimise the use of jargon and abbreviations unless intuitive explanatory mechanisms, such as tooltips or training are provided. For example, the WCAG success criteria 3.1.3 (Unusual Words) and 3.1.4 (Abbreviations) \cite{wcag2.2} require that users be given accessible definitions for uncommon terms, idioms, and abbreviations. 

\subsection{APD 6: Fear of Making Mistakes in Apps} \label{fear of mistakes}

\begin{figure}
\includegraphics[width=\textwidth]{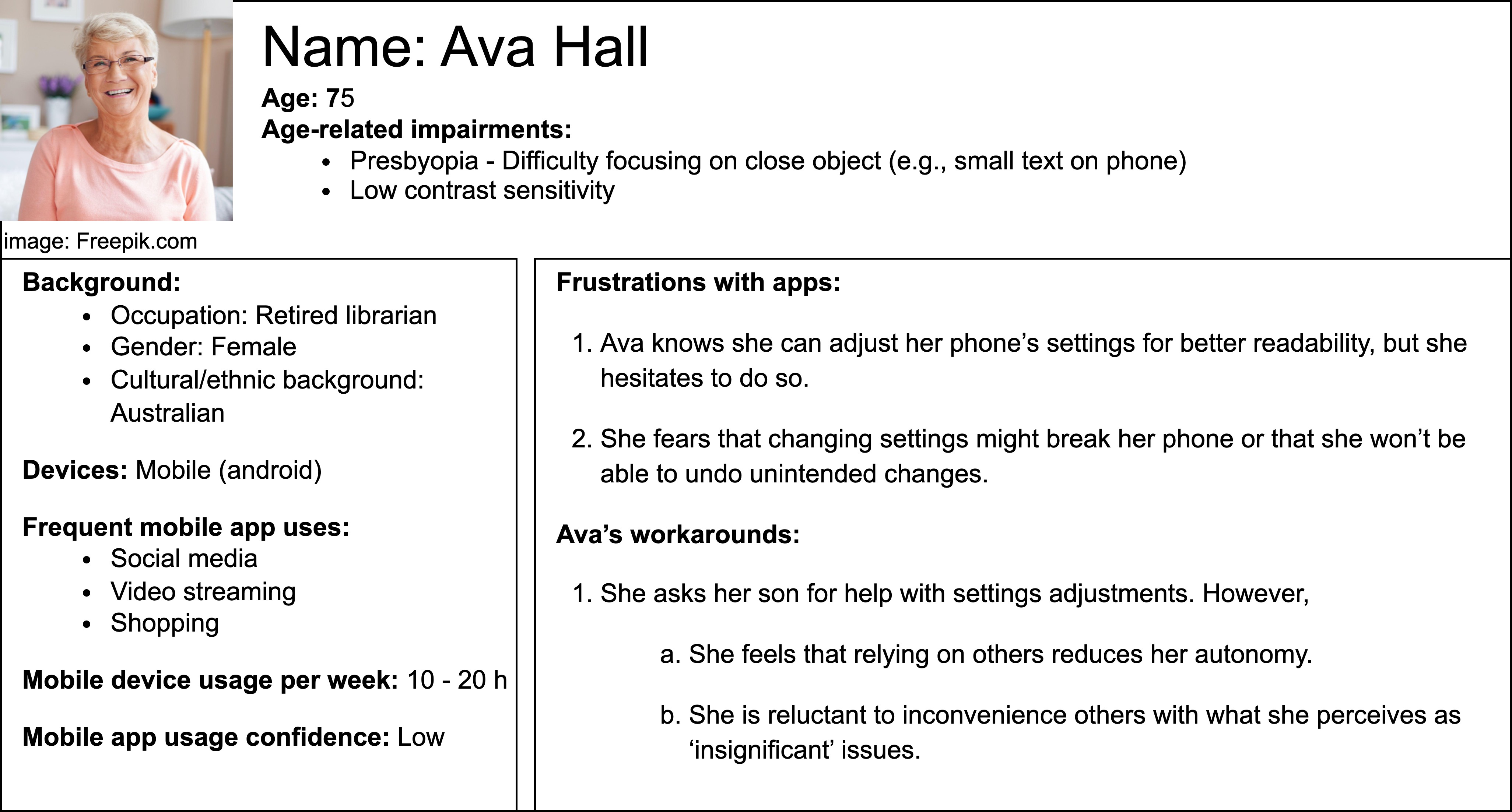}
\caption{A persona representing a fictional senior user, Ava, representing how seniors fear making changes in app/device personalisation settings due to their unfamiliarity with the technology}\label{ava_persona}
\Description{The figure presents a persona of senior user named Ava Hall who has low vision issues. Her impairments coupled with her lack of confidence in using technology, has made her fearful to make any changes in her apps or the mobile device to change any accessibility setting to make her app experience better. She would ask for help from people who are close to her when she perceives that the request is not causing any inconveniences for others.}
\end{figure}

\subsubsection{\textbf{Problem:}}

Let us consider Ava's persona depicted in Figure \ref{ava_persona}. Her reading experience in apps is affected by her vision limitations, which have gradually developed with age. She has observed her friends at the local U3A chapter using larger text sizes and enhanced colours on their phones. However, she hesitates to navigate her phone's settings to make similar adjustments, as she is unfamiliar with the available options and fears that pressing the wrong button might break her phone. 
Ava is also reluctant to ask her friends for assistance, as they themselves are hesitant to modify another person's phone due to concerns that any unintended changes could lead to negative consequences for a device they do not own. As a result, if she wishes to adjust her phone settings, she relies on her son or grandchildren for help. Nevertheless, she perceives her request as trivial and worries about inconveniencing others. Her struggle to maintain autonomy in the face of age-related limitations is further challenged by her frequent need to seek assistance with regard to device and app usage.  

This fear was a recurring theme across both the exploratory and evaluative phases of the study (FG-A, FG-B, FG-C, FG-D, and FG-E). It was particularly evident among seniors who lacked confidence in their IT skills, often manifesting as hesitation or anxiety when attempting to modify app settings or functionalities. Several factors appeared to contribute to this fear, including the widespread targeting of seniors by predatory scams (FG-B), general unfamiliarity with computing devices and software (FG-C and FG-D), and a perceived lack of competence in operating computing software and hardware (FG-C, FG-D, and FG-E). Unlike younger generations, many seniors were not raised in a technology-driven environment, making it more challenging for them to adapt to app user interfaces. For example, the following statement from a participant in FG-C illustrates this dilemma:

\begin{quote}
\textit{"Because they [seniors] are too scared that if they do it [touch a button], something's going to happen to their phone. Whereas a person who is used to technology will touch it [button], but they'll say, I'm still going to hurt my phone, [but] I'm just going to play around with it [anyways], and that is the big difference [between seniors and younger generations]."}
\end{quote}

However, our discussions with seniors also revealed several examples across both focus groups of seniors who were willing to experiment and familiarize themselves with apps and devices to improve their digital proficiency. According to some participants from FG-C, FG-D, and FG-E, it isn't just enough to provide seniors with guidance. We, as developers, should facilitate a safe, consequence-free environment in the apps we develop for trial and error experimentation by seniors. It improves learning and confidence in navigating app interfaces. 

\subsubsection{\textbf{Recommendations:}}

The fear or reluctance that many older adults experience when using applications, particularly when adjusting app- or device- level accessibility configurations to personalise their user interfaces, is a well-documented phenomenon in existing literature \cite{wcag2.2, iso9241, iosDevGuidelines, inostroza2016, salman2018, Shamsujjoha2025, harte2017, watkins2014, holden2020, barbosa2015, conte2019, almeida2015, holzl2013, morey2019, harte2018}. As this issue was not addressed in the user study conducted with our current prototype iteration, our recommendations are grounded in the findings of prior studies.

We  \textbf{\textit{recommend that developers always incorporate \textit{undo} or \textit{abort} mechanisms}} to allow users to revert recent activations or deactivations of accessibility features or other app-level configuration changes \cite{iso9241, inostroza2016, Shamsujjoha2025, watkins2014, holden2020}. This functionality \textbf{\textit{should be complemented by a ‘reset' mechanism or an ‘emergency exit' that restores the application to a default state (or a desired state) in case the user believes an error has been made}} during personalisation \cite{inostroza2016}. To avoid complete loss of personalisation progress, developers could \textbf{\textit{implement a feature to save customisation profiles within the application}}. For instance, appearance settings such as font size and colour themes could be saved for future reuse \cite{iso9241}.

Additionally, we  \textbf{\textit{recommend the integration of in-app help mechanisms to support older users}}. Basic accessibility features could be accompanied by short, actionable tooltips, pop-ups, or hints to explain their function \cite{iosDevGuidelines, morey2019, salman2018, conte2019, almeida2015}. For more complex features, tooltips might be enhanced with short animations demonstrating the outcome of a change or preview options before confirmation \cite{iosDevGuidelines}. These aids can be particularly beneficial for older users experimenting with configuration changes for the first time \cite{iosDevGuidelines, morey2019, holzl2013}.
Easy access to clear, concise guidance materials and training resources can foster confidence among older users when interacting with applications \cite{iso9241, iosDevGuidelines, inostroza2016, Shamsujjoha2025, barbosa2015, morey2019}. However, it is essential that all help content avoids unfamiliar terminology, including technical, cultural, and generational jargon, as well as ambiguous terms, idioms, and abbreviations.

\subsection{APD 7: Getting Lost Within Apps (When Navigating)} \label{getting lost}

\begin{figure}
\includegraphics[width=\textwidth]{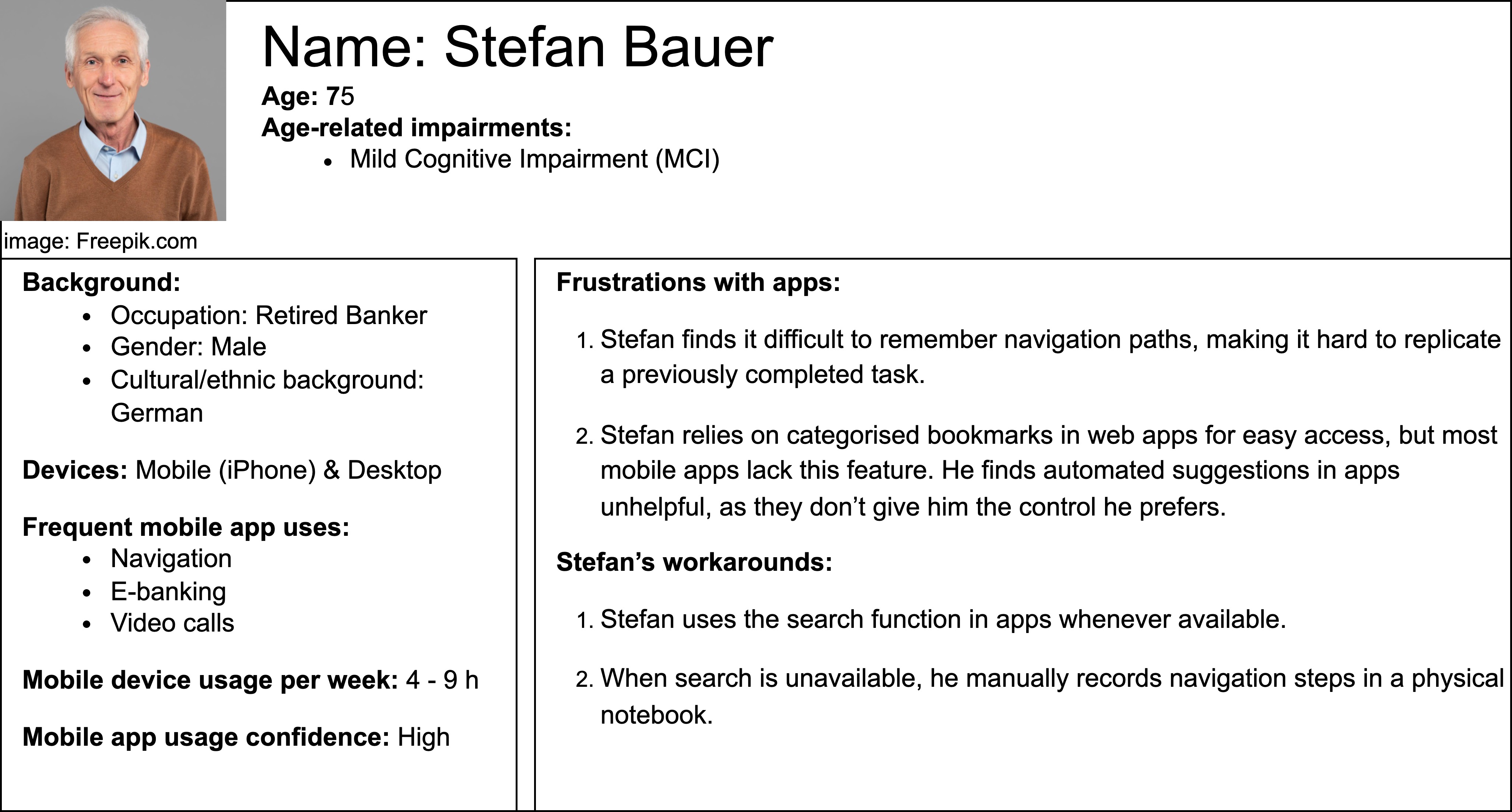}
\caption{A persona representing a fictional senior user, Stefan, who due to his age-related cognitive decline, is getting lost within the menus of apps.}\label{stefan_persona}
\Description{A fictional persona representing a senior user named Stefan, who is affected by with Mild Cognitive Impairment. Memory related issues caused by these memory impairments are causing him to forget navigational pathways in apps, which has led him to either rely or search functions in apps when available or using a physical notebook to write down navigation steps.}
\end{figure}

\subsubsection{\textbf{Problem:}}

In Figure~\ref{stefan_persona}, we illustrate the challenges faced by seniors experiencing age-related cognitive decline through the persona of Stefan. Stefan lives with Mild Cognitive Impairment (MCI), which affects his memory in routine activities, such as remembering appointments and recalling past events or conversations. This condition similarly impairs his ability to memorise app navigation pathways. This issue reflects a recurring theme identified during our problem exploration phase (FG-A and FG-B): seniors frequently become disoriented while navigating app menus and pages. 

A strategy Stefan effectively uses in web apps is bookmarking pages he frequently revisits. He also manages a large collection of bookmarks by categorizing them within his browser. However, this method is rarely available in mobile apps. For example, when using a retail app, seniors like Stefan struggle to retrace their steps to rediscover a specific product. One participant from FG-A described resorting to pen and paper as a workaround:  

\begin{quote}
\textit{"But in some of them [apps], they don't have a bookmark because you're not actually on the general internet search thing. You're within a program or within a site. So all you can do is, you have to remember by pen and paper where you went and what you had to do and it's not always that easy."}
\end{quote}  

For Stefan, the most effective way to avoid getting lost within apps is to use the search function whenever available. Many seniors expressed a preference for search over manual navigation in both exploratory stage focus groups (FG-A and FG-B).

\subsubsection{\textbf{Recommendations}}

To address the navigation difficulties often experienced by older users, particularly the tendency to become disoriented within application pages and menus, we propose several recommendations. Based on insights gathered from our exploratory focus groups, we first \textbf{\textit{recommend the inclusion of a ‘search' feature prominently on the application's landing page}}. This enables users to directly locate desired content or features without navigating through multiple menus. Secondly, \textbf{\textit{we recommend implementing a bookmarking mechanism}} that allows users to save specific app pages for easy revisitation. This is particularly beneficial in applications with numerous pages, such as those in retail, social media, or streaming domains.

We also draw upon existing literature to support additional recommendations for improving navigational clarity for older users. A key recommendation is \textbf{\textit{the provision of a ‘safe space'}}: a clearly identifiable home or landing page that users can return to when they feel disoriented. The icon or button linking to this page should be prominently displayed within the user interface to ensure easy access \cite{redeap, almeida2015, barros2014, morey2019}. Equally important is the \textbf{\textit{inclusion of a consistently visible ‘back' button}}. This button should reliably return users to the previous screen and allow them to retrace their navigational steps in a linear and predictable manner \cite{redeap, barros2014}. Repeated use of this button should take the user backwards along their navigational pathway, ultimately returning them to a familiar and stable point within the application. To further enhance navigational awareness, we recommend \textbf{\textit{incorporating a mechanism to clearly indicate the user's current location}} within the app. Techniques such as \textit{breadcrumb trails} can serve this purpose effectively by showing the hierarchical structure of the user's navigational path \cite{Shamsujjoha2025}.

\subsection{APD 8.1: Audio Modality -- Need for Voice Input} \label{voice input}

\subsubsection{\textbf{Problem:}}

To illustrate this issue, we revisit the persona of Liam, presented in Figure~\ref{liam_persona}. Liam experiences difficulties when using his phone’s virtual keyboard due to limited finger dexterity and joint stiffness. Additionally, his fingers are often swollen and dry, which significantly hinders his ability to perform the precise touch actions required for virtual keyboard input. A similar condition is described in the persona of Carl (Figure~\ref{persona example}), whose practical workaround involves the use of a third-party stylus to improve typing accuracy. These mobility-related impairments were discussed throughout both the exploratory and evaluative phases of the study (FG-A, FG-B, FG-C, FG-D, and FG-E). Notably, at the start of an evaluative focus group (FG-E), one participant proactively asked whether we had considered integrating an audio input modality -- prior to us even suggesting such a feature -- highlighting the relevance and urgency of addressing this accessibility need.

\begin{quote}
    \textit{Is there, maybe for people that still can see but have arthritis, an audio aspect, like you sometimes have in on your search [fields], [...] But is there something like that [voice input] has been thought of, or there's not?}
\end{quote}

\subsubsection{\textbf{Our Solutions:}}

\begin{figure}
\includegraphics[width=0.6\textwidth]{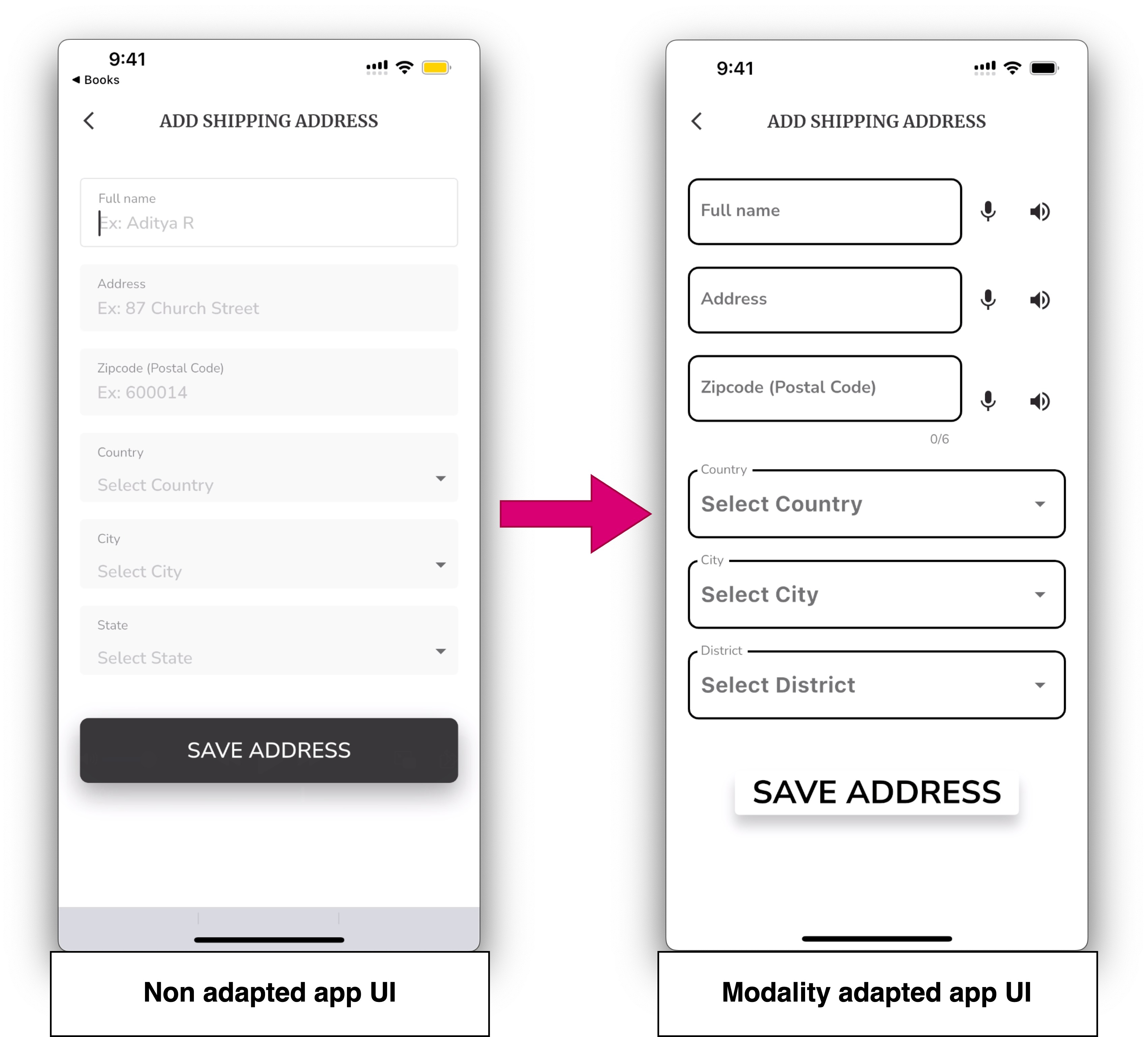}
\caption{Modality adapted UI example. In here, the first three text input fields in the non-adapted UI are transformed into widgets that have text-to-speech and speech-to-text features embedded in them.}\label{modality ui}
\Description{This figure presents a comparison between two mobile app user interfaces. The left interface displays a standard form with several input fields. In the right interface, the first three input fields have been enhanced with speech-to-text and text-to-speech functionalities.}
\end{figure}

Participants from all three evaluative focus groups (FG-C, FG-D, and FG-E) reacted positively to the speech-to-text feature enhancements embedded in the adaptive app prototype (Figure~\ref{modality ui}). In this example, we integrated both text-to-speech and speech-to-text functionality for each input field in a form and asked seniors whether such inclusion was beneficial to them. The following response from FG-E illustrates this perspective:  

\begin{quote}
\textit{"Well with Siri and all these things [voice assistants], it's just common sense that these things [text-to-speech and speech-to-text features] should be available whenever it's [app] asking you for information. You've got to be able to speak to it. Now, typing and having to read..., that is a thing for the younger people."}
\end{quote}  

\subsubsection{\textbf{Recommendations:}}

Among the existing accessibility standards, ISO 9241 \textit{Ergonomics of human-system interaction}, Part 171: \textit{Guidance on software accessibility}, strongly recommends that developers provide speech recognition services in software where the user’s hardware supports this capability \cite{iso9241}. We concur with this recommendation and \textbf{\textit{we argue that speech recognition functionality is essential for applications designed for older users}}. Several other studies corroborate this perspective \cite{Shamsujjoha2025, mi014, barros2014}.
We also recommend \textbf{\textit{the integration of voice input features at the application level}}, rather than relying solely on the speech-to-text functionalities native to the operating system. Our focus group sessions revealed that many older users are either unaware of or do not actively use the OS-native voice input features. Embedding voice input functionality directly within the application, such as by including a voice input button within an input field widget (as illustrated in Figure~\ref{modality ui}), can enhance both accessibility and intuitiveness.

As noted by Shamsujjoha et al. \cite{Shamsujjoha2025}, users should have the ability to toggle voice input capabilities during both the installation/setup phase and throughout active usage of the application. Thus, we also \textbf{\textit{recommend voice input be a user-controlled optional feature}}, particularly in cases where developers intend to cater to both senior users and younger generations who may not require voice input functionality \cite{wickramathilaka2025}.

\subsection{APD 8.2: Audio Modality -- Need for Screen Reading} \label{screen reader}

\begin{figure}
\includegraphics[width=\textwidth]{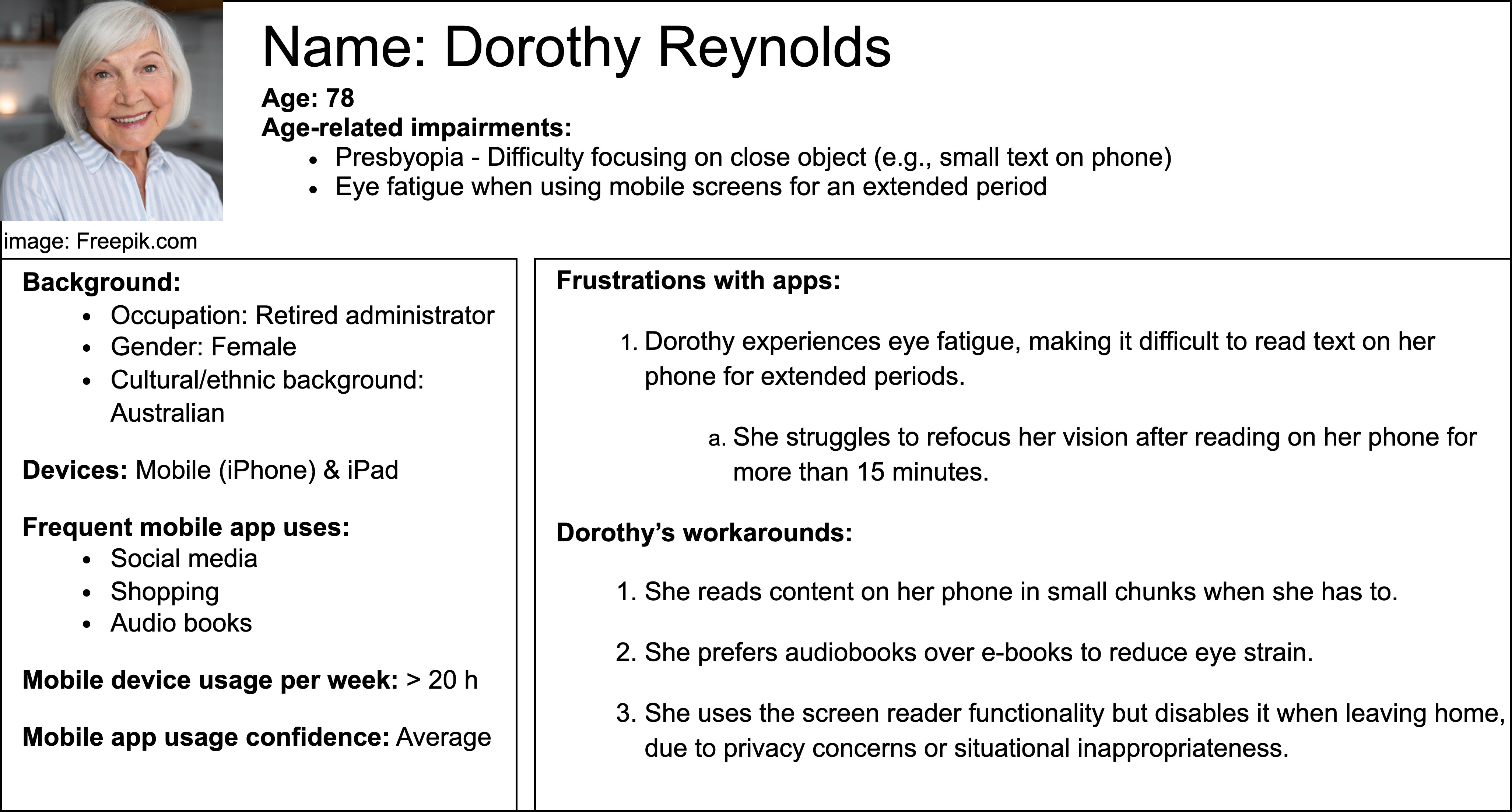}
\caption{A persona representing a fictional senior user, Dorothy, who finds it difficult to read anything on her phone for an extended time.}\label{dorothy_persona}
\Description{A fictional persona of a senior named Dorothy has been represented in this figure. She has eye-fatigue when attempting read anything on her mobile device. She is pivoting more towards audio modality through audiobooks and screen reader feature due to this issue.}
\end{figure}

\subsubsection{\textbf{Problem:}}

In Figure \ref{dorothy_persona}, we illustrate the frustrations of Dorothy, who struggles to focus on her phone for extended periods. If she persists despite accumulating eye fatigue for more than 15 minutes, she experiences difficulty refocusing on distant objects for approximately half a minute after looking away from the screen. This issue was reported by a participant during one of our exploratory focus groups (FG-B) with the following statement:  

\begin{quote}
\textit{"It's the focus. As you get older, your [eye] focus takes time."}
\end{quote}  

Besides eye strain, many other age-related vision impairments may impact the reading experience of seniors as illustrated by personas such as Kathryn (Figure \ref{kathryn_persona}) and Judy (Figure \ref{judy_persona}). A workaround Dorothy found in particular is to read news articles, social media content, or e-books in short intervals. A different solution is switching to an alternative output modality, such as screen readers or audiobooks.   

During our exploratory stage (FG-A and FG-B), we asked participants whether they would prefer apps to read articles or content aloud to gauge their interest in screen-reading features. Both focus groups agreed that such functionality would be beneficial for accessibility.  

However, integrating these features introduces situational considerations. A recurring theme across all focus groups was the appropriateness of using audio input and output in public spaces. Many seniors expressed a strong reluctance to use audio in such settings due to social norms. For example, the following statement was made in FG-A:  

\begin{quote}
\textit{"I was in the shopping centre this morning. I looked into two conversations in the toilet [on their phones], thinking, oh, how ridiculous. No..., put your earphones in. I don't need people hearing what I'm listening to."}
\end{quote}

Privacy concerns were another major factor influencing participants' hesitation to use these features in public settings. The following example was taken from FG-C:  

\begin{quote}
\textit{"Everybody would know everything... my address, personal details and even [what I] just wanted to order."}
\end{quote}  

Nonetheless, participants were open to using audio-based features to improve their app experience. For example, a participant from FG-B noted that they would use these features with earphones on, both for privacy and to enhance audibility in noisy environments (FG-B). Most seniors would opt for the simplest solution: not using the screen reader functionality in public situations (FG-D and FG-E). 

\subsubsection{\textbf{Our Solution}}

During the evaluation phase, we showed our participants examples where text-to-speech functionality was integrated into an input field and shown to end users via a button with a microphone icon (Figure \ref{modality ui}). Participants across all three evaluative focus groups (FG-C, FG-D, FG-E) appreciate our efforts to provide an alternative audio modality in general and specific mentions about text-to-speech/screen reading functionality were found in 2 out of the 3 sessions (FG-D and FG-E). For example, the following statement was made by one of the seniors who participated in FG-D:

\begin{quote}
\textit{"Because, to me, it is because of eyesight, especially when you're on a small screen. So anything that's coming out audible -- obviously, we're going to have trouble with our hearing too over time, I understand that -- but because, you know, it might say the price [of an item]. Because now, when something is really small, I have to go open over and over [with zoom gestures] to make it bigger, bigger, bigger so I can actually read. [...] Because [having audio modality in apps], it's just a start. But, something audible would be very helpful for me."}
\end{quote}

\subsubsection{\textbf{Recommendations}}

Both the Apple Developer Accessibility Guidelines \cite{iosDevGuidelines} and Android Developer Guidelines \cite{androidDevGuidelines} provide guidance on integrating the respective platform’s native screen-reading functionality. However, \textbf{\textit{we recommend that developers implement text-to-speech functionality at the application level}}. Similar to voice input, our findings indicate that many older users are either unaware of or do not actively use the screen-reading features provided natively by the operating system. Consequently, embedding a more visible and easily accessible screen-reading feature directly within the application interface, as demonstrated in Figure~\ref{modality ui}, can significantly enhance accessibility, particularly in situations where users prefer auditory feedback over visual interaction due to environmental or physical constraints.
This recommendation is supported by previous studies, including those by Shamsujjoha et al. \cite{Shamsujjoha2025} and Mi et al. \cite{mi014}, both of which emphasise the necessity for app-level screen-reading functionality. ISO 9241-171 also advocates for enabling such features wherever technically feasible \cite{iso9241}.

An important \textbf{\textit{complementary recommendation is to ensure that sensitive information is excluded from being read aloud}} by either app-level or OS-level screen readers. Developers should take explicit measures to exclude private data, such as financial details (e.g., bank balances and account numbers), authentication credentials (e.g., passwords and social security numbers), personal health data, and personal contact details, from screen reader accessibility. This precaution helps to prevent inadvertent disclosure in situations where the user is not in a private or secure environment. A more advanced approach could involve \textbf{\textit{restricting screen-reading functionality to predefined, geo-fenced safe locations identified by the user}}.

\subsection{APD 8.3: Audio Modality -- Validating Voice Input Through Screen Reading}

\subsubsection{\textbf{Problem and Our Solution}}

even though participants expressed a positive attitude toward switching to vocal modalities when necessary, we observed a general scepticism across all three evaluative focus groups (FG-C, FG-D, and FG-E) regarding senior confidence in using speech-to-text functionality to input accurate information into the app. This scepticism was based on their own experiences and their empathy for others who may have diverse accents and intonations when speaking English. Inaccurate voice inputs may be more time-consuming task to a set of seniors due to their speech characteristics to a point where it would be more efficient to use the on-screen keyboard, despite its inherent accessibility barriers. One participant from FG-D illustrated this point:  

\begin{quote}
\textit{"It [voice input] could end up being more time-consuming if it [speech-to-text feature] continues to get it wrong."}
\end{quote}

But an unexpected insight that we identified from 2 out of 3 evaluative focus groups (FG-C and FG-D) was regarding how well received the simultaneous integration of speech-to-text and text-to-speech features was as using them in tandem appears to mitigate some of the limitations of voice input technology. For example, if Liam (Figure \ref{liam_persona}) is having issues with inaccurate voice inputs, then being able to first, input the text through speech and then use a screen reader functionality to read out loud the content that he just input in the app, enables him to validate his voice inputs. 

\subsubsection{\textbf{Recommendations}}

Guidelines proposed in several studies \cite{redeap, Shamsujjoha2025, mi014} have highlighted the importance of enabling users to validate app inputs through auditory feedback. We believe this approach can effectively address the issue of inaccurate voice inputs observed among our senior user study participants. As such, \textbf{\textit{we recommend that developers provide senior users with the ability to toggle both speech-to-text and text-to-speech features concurrently}} when entering information into the application (e.g., as illustrated in Figure~\ref{modality ui}).
For example, when inputting an address, a senior user could first utilise the speech-to-text functionality to provide the input vocally. Subsequently, the screen-reading feature can be used to read the input data back to the user, allowing them to confirm its accuracy. This bidirectional audio interaction not only improves input validation but also enhances overall confidence and usability for senior users.

\subsection{APD 9.1: Personalisation -- Too Much Complexity}

\subsubsection{\textbf{Problem}}

A recurring theme across all three of our evaluative focus groups (FG-C, FG-D, and FG-E) was the concern that introducing a personalisation configuration layer into an application could significantly increase complexity for many senior users. For instance, when we proposed a scenario in which seniors could issue natural language commands, either vocally or textually, to personalise the application through a multimodal Large Language Model (LLM), participants expressed considerable scepticism regarding its usefulness. In both FG-D and FG-E, participants perceived this functionality as overly complex and of little value to them. Furthermore, these participants reported uncertainty about how to articulate natural language commands to achieve their desired personalisation outcomes. As one participant noted:

\begin{quote}
\textit{"But I also think that for people my mother's age, which is 90, she wouldn't even know what to ask. Okay, so 'change the colour button' is fairly straightforward, but anything more complex than that, she wouldn't even know how to go about phrasing or what words to use."}
\end{quote}

\subsubsection{\textbf{Recommendations}}

Across all three evaluative groups (FG-C, FG-D, and FG-E), the most preferred way of applying app run-time adaptations seems to be a configuration dashboard that is simple to use. This became evident to us when, among all three proposed personalisation methods, it made the most sense to our participants. The following example from FG-C is a good indicator of this sentiment:

\begin{quote}
\textit{"If you're talking about just this settings screen, just keeping it as simple [as possible]. Most people aren't going to need all those [adaptation options]. An older person isn't going to worry about changing all this and [so,] keep it very, very simple, the options."}
\end{quote}

Therefore, we \textbf{\textit{recommend that developers provide accessibility app configurations primarily through a traditional configuration panel}}, rather than relying on self-adaptation or Large Language Model (LLM)-driven techniques. However, these configuration panels can still be too complex for seniors and therefore, we \textbf{\textit{recommend that the configuration panel should include only a limited set of essential personalisation options (e.g., enhancing contrast, increasing readability, switching themes, and changing modalities)}}. Some seniors with higher confidence, skills, and familiarity in their technology usage may also be benefitted by an \textbf{\textit{advanced configuration panel that offers more detailed personalisation options}}.

\subsection{APD 9.2: Personalisation -- Lack of Confidence in Technological Skills}

\subsubsection{\textbf{Problem}}

Another distinct theme that emerged from the data across all three evaluative focus groups (FG-C, FG-D, and FG-E) was the hesitation among senior participants to use personalisation features due to low confidence in using technology and mobile applications. One participant, who was personally confident in their technological skills and willing to experiment with accessibility settings, highlighted a contrasting experience among their peers:

\begin{quote}
\textit{"I know a lot of older people are scared of doing it [changing setting], so they're not. We're familiar with it because we've sort of grown up with it [technology] a bit. But a lot of my older friends, they wouldn't use it because of the complexity and the fact that they haven't got the confidence to use it."}
\end{quote}

This lack of confidence often stemmed from a fear of making mistakes during configuration changes, which we also explored under APD 6: Fear of Making Mistakes in Apps (Section \ref{fear of mistakes}). In some cases, this issue appeared to be exacerbated by unintuitive design choices made by developers in presenting accessibility configuration options. Such designs could result in unexpected behaviour when a setting is changed or could make it difficult for seniors to even locate accessibility settings. Both of these scenarios were observed during our study. The following quote from a participant in FG-C illustrates the latter case:

\begin{quote}
\textit{"I have to say, I've never even pressed on that [accessibility settings option on a mobile device] to see what it does. So maybe because I haven't needed to. But yeah, if it was at the top, I probably would look at it more readily."}
\end{quote}

\subsubsection{\textbf{Recommendations}}

One of the key ideas discussed across all three evaluative focus groups (FG-C, FG-D, and FG-E) was that, regardless of the form in which run-time app personalisation functionality is presented to seniors, it is vital that they receive clear instructions and guidance on how to operate these adaptations. For example, the following statement was made by a participant in FG-D:

\begin{quote}
\textit{"I think because once the trust is there, and [if] it is in language that we're used to, [communicated in] the way we learn in our generation, then we can familiarize ourselves [with personalisation methods], and then it becomes just like walking, you know? All of a sudden we know just to press the accessibility button for that particular app"}
\end{quote}

In the first place, access to app-level accessibility configurations should be convenient for seniors. Therefore, we \textbf{\textit{recommend that any such configuration options should be accessible from all app interfaces through a prominently placed ‘Settings' button}} that does not obstruct other UI elements. \textbf{\textit{A tooltip or a similar guiding mechanism should be in place to clearly convey the purpose of the said button}}.

Inside these configuration panels, we recommend that each personalisation option be accompanied by a clear, jargon-free explanation (e.g., via a tooltip), and a short preview of the outcome of the configuration change (similar to the APD 6.d recommendation). These configurations should also be supported by short, in-app tutorials that can further assist seniors in understanding these options.

The above recommendations are especially pertinent when developers intend to use multi-modal LLMs to accept both text and voice prompts from seniors to trigger app personalisation. In such cases, we recommend that \textbf{\textit{example natural language prompts on how to request app adaptations be provided for seniors}} to better understand how to structure their requests.

\subsection{APD 9.3: Personalisation -- Reduction of Autonomy}

\subsubsection{\textbf{Problem}}

One of the goals of our study was to identify a personalisation strategy that would be readily adopted by senior users. However, when we proposed the concept of app self-adaptation as a possible strategy, participants expressed strong negative reactions. A dominant theme across all evaluative focus groups (FG-C, FG-D, and FG-E) was that such approaches were perceived to undermine seniors’ sense of control and autonomy. The following statement from a participant in FG-C encapsulates this sentiment:

\begin{quote}
\textit{"I find it [self-adaptive apps] annoying, sorry. I would hate it because I feel I've lost control."}
\end{quote}

A key consequence of seniors' desire to maintain control over their applications and personalisation settings is their reluctance to accept automatic adaptations, particularly when the changes are not clearly explained. This need for transparency and understanding was emphasised in discussions from both FG-C and FG-E. 

\subsubsection{\textbf{Recommendations}}

To preserve seniors' sense of control and autonomy, we \textbf{\textit{recommend that developers choose user-controlled adaptations over self-adaptive apps}}. Among the personalisation strategies discussed, self-adaptive applications elicited the strongest resistance from participants, particularly in comparison to traditional configuration panels and multi-modal LLM prompt-based approaches. If, for any reason, self-adaptation features are to be implemented, \textbf{\textit{they must not compromise the user's sense of control and should always request permission before proceeding}}.

\subsection{APD 9.4: Personalisation -- Feeling of Intrusiveness}

\subsubsection{\textbf{Problem}}

Another crucial concern raised in two of the evaluative focus groups (FG-D and FG-E) was the perceived intrusiveness of self-adaptive applications. When we introduced the idea of an app that could adapt itself based on user behaviour and parameters, many participants expressed discomfort, reinforcing the previously noted issue of diminished autonomy. Several seniors felt that such applications would be overly invasive. For instance, during one discussion, participants drew parallels between self-adaptive apps and the personalised advertisements they encounter on social media platforms after having mentioned a product aloud. 

\subsubsection{\textbf{Recommendations}}

Seniors indicated that self-adaptive applications in particular, can evoke a sense of intrusiveness. We believe that opting for alternative personalisation mechanisms, such as traditional configuration panels, presents a much lower risk of eliciting such perceptions among senior users. In cases where developers intend to implement self-adaptive features that automatically respond to changes in a senior user's contextual parameters, these \textbf{\textit{features should be designed in a manner that does not compromise the user’s sense of privacy}}.

One effective method to mitigate this risk is to \textbf{\textit{foster trust between the user and the application by providing a clear, upfront explanation of how and why a self-adaptation was triggered}}. Additionally, \textbf{\textit{users should be given the ability to undo an app's self adaptive action}}. A participant from FG-C illustrated how seniors may be more receptive to self-adaptive behaviour when they are able to understand what is occurring:

\begin{quote}
\textit{"I would like it, if I understand it. [For example,] if there's something that my grandson has done with it [app/device]: ‘Look quickly, you do it like this.' I think, 'What has he done?' But I just accept it [configuration change]. [When] automatic aspects have [been] changed automatically, I [need to] understand that they've changed automatically, and my grandson hasn't changed it."}
\end{quote}

\subsection{Summary of Key Recommendations} \label{recommendations} 

{
    \renewcommand{\arraystretch}{1.3}
    \scriptsize
    
    \begin{longtable}{|>{\raggedright\arraybackslash}p{2cm}|
                      >{\raggedright\arraybackslash}p{1cm}|
                      >{\raggedright\arraybackslash}p{9.5cm}|}
    
        \caption{A summary of accessibility and personalisation recommendations} \label{tab:accessibility_recommendations} \\
        \hline
        \textbf{Accessibility Problem Dimension (APD)} & \textbf{ID} & \textbf{Accessibility Recommendations} \\
        \hline \hline
        \endfirsthead
        
        \hline
        \textbf{Accessibility Problem Dimension (APD)} & \textbf{ID} & \textbf{Accessibility Recommendations} \\
        \hline \hline
        \endhead
        
        \hline
        \endfoot
        
        \hline
        \endlastfoot
    
            \multirow{6}{=}{APD 1: Enhancing text readability} & APD 1.a & Use a minimum font size of 20 pt for body text in mobile applications targeting seniors. Whenever possible, strive to keep all text in the app above the threshold.\\ \cline{2-3}
             & APD 1.b & Use a minimum font weight of 700.\\ \cline{2-3}
             & APD 1.c & If a gesture-based zoom feature must be implemented, use a magnifying glass widget or overlay instead.\\ \cline{2-3}
             & APD 1.d & Use serif and sans serif typefaces such as Helvetica, Arial, and Times New Roman. If the development platform recommends system font families (e.g., iOS or Android), consider using those.\\ \cline{2-3}
             & APD 1.e & Ensure that font size adjustments made by the user (via app-level or OS-level configurations) do not cause UI misalignments or content truncation.\\ \cline{2-3}
             & APD 1.f & Maintain line spacing of at least 1.5 times the font size , spacing between paragraphs be at least 2 times the font size, letter spacing be at least 0.12 times the font size, and word spacing of at least 0.16 times the font size.\\ \hline \hline
             
             APD 2: Enhancing contrast and use of colour themes & APD 2.a & Implement both a 'light' theme (white background, black foreground) and a 'dark' theme (black background, white foreground), aiming for the maximum possible contrast ratio (21:1).\\ 
             & APD 2.b & Where monochrome is not suitable, colour may be used, provided it maintains high contrast. Use at least a 7:1 contrast ratio for small/body text, and 4.5:1 for larger text (e.g., headings).\\ \cline{2-3}
             & APD 2.c & Allow users to select from a predefined set of high-contrast themes tailored to common impairments such as macular degeneration, colour blindness, and presbyopia.\\ \cline{2-3}
             & APD 2.d & Enable users to create custom themes using a restricted colour picker that enforces minimum contrast ratios.\\ \cline{2-3}
             & APD 2.e & Implement a theme management and propagation mechanism from your development framework (e.g., Flutter’s \textit{Theme} widget).\\ \cline{2-3}
             & APD 2.f & Evaluate themes under varying lighting conditions and hardware features such as Apple’s True Tone to ensure contrast consistency.\\ \hline \hline
             
             \multirow{2}{=}{APD 3: Enhancing touch targets} & APD 3.a & Ensure a minimum touch target size of 7--10 mm in both dimensions. Use larger targets where layout permits.\\ \cline{2-3}
             & APD 3.b & Maintain at least 8 dp of spacing between adjacent touch targets.\\ \hline \hline
             
             \multirow{3}{=}{APD 4: Mitigating limitations of mobile screen size} & APD 4.a & Use a wizard pattern to segment and reflow large content sections, reducing vertical scrolling caused by enlarged text.\\ \cline{2-3}
             & APD 4.b & Apply the wizard pattern to break long forms (e.g., user registration) into smaller, manageable steps to reduce cognitive overload.\\ \cline{2-3}
             & APD 4.c & Minimise the use of on-screen virtual keyboards where feasible.\\ \hline \hline
             
             \multirow{4}{=}{APD 5: Error handling, language, and jargon} & APD 5.a & Proactively prevent errors via rigorous QA to retain senior users.\\ \cline{2-3}
             & APD 5.b & When errors occur, communicate them in layperson-friendly language. Avoid technical, cultural, or generational jargon, idioms, humour, and abbreviations.\\ \cline{2-3}
             & APD 5.c & Provide clear, step-by-step guidance to help seniors independently resolve errors.\\ \cline{2-3}
             & APD 5.d & Ensure that help content and access to support personnel are easily available and jargon-free.\\ \hline \hline
    
             \multirow{4}{=}{APD 6: Mitigating seniors’ fear of making mistakes} & APD 6.a & Provide an \textit{undo} mechanism to revert recent changes to app accessibility settings.\\ \cline{2-3}
             & APD 6.b & Provide a \textit{reset} or an \textit{emergency exit} mechanism to return the app to a default or a desired state.\\ \cline{2-3}
             & APD 6.c & Allow users to save personalisation profiles (e.g., custom themes).\\ \cline{2-3}
             & APD 6.d & Offer in-app guidance using tooltips, pop-ups, or hints to explain functions and configurations.\\ \hline \hline
    
             \multirow{5}{=}{APD 7: Supporting app navigation experience} & APD 7.a & Include a \textit{search} feature as a primary navigation method.\\ \cline{2-3}
             & APD 7.b & Provide bookmarking functionality for saving or favouriting frequently visited app pages, particularly in multi-page apps (e.g., retail, streaming).\\ \cline{2-3}
             & APD 7.c & Include a clearly visible home or landing page button on all pages to return users to a familiar safe space.\\ \cline{2-3}
             & APD 7.d & Ensure a prominent \textit{back} button is available (except on the home page) to allow consistent backward navigation.\\ \cline{2-3}
             & APD 7.e & Display the user's current location using a \textit{breadcrumb trail}.\\ \hline \hline
    
             \multirow{5}{=}{APD 8: Supporting audio modality} & APD 8.a & Allow users to toggle voice input and screen reader functionality at both setup and runtime.\\ \cline{2-3}
             & APD 8.b & Implement audio modality (voice input \& screen reader) at the app level, enhancing specific UI elements such as input fields and text blocks.\\ \cline{2-3}
             & APD 8.c & Exclude sensitive or private information (e.g., passwords, account numbers) from screen reader output to prevent accidental disclosure.\\ \cline{2-3}
             & APD 8.d & Provide automated toggling of audio modality based on predefined safe spaces (e.g., designated home or geo-fenced area).\\ \cline{2-3}
             & APD 8.e & Support tandem use of voice input and screen reading. For example, to input data by voice and confirm it using audio feedback.\\ \hline \hline

             \multicolumn{3}{|l|}{\textbf{Tentative app personalisation recommendations}} \\ \hline

             \multirow{10}{=}{\parbox{2cm}{\raggedright APD 9: Supporting app personalisation}} & APD 9.a & Accessibility app configurations should be primarily provided through a traditional configuration panel.\\ \cline{2-3}
             & APD 9.b & Initially, the configuration panel should include only a limited set of essential personalisation options (e.g., enhancing contrast, increasing readability, switching themes, and changing modalities).\\ \cline{2-3}
             & APD 9.c & Seniors who are more confident or technically inclined should be able to access an advanced configuration panel that offers more detailed personalisation options.\\ \cline{2-3}
             & APD 9.d & Accessibility configuration options should be accessible from all app interfaces through a prominently placed ‘Settings' button that does not obstruct other UI elements.\\ \cline{2-3}
             & APD 9.e & A tooltip or a similar guiding mechanism should be in place to clearly convey the purpose of the said ‘Settings' button\\ \cline{2-3}
             & APD 9.f & If adaptations are to be driven by a multi-modal Large language Model, ensure the feature is supported with example natural language prompts on how to request app adaptations.\\ \cline{2-3}
             & APD 9.g & Opt for user-controlled adaptations over the self-adaptation of apps.\\ \cline{2-3}
             & APD 9.h & If any self-adaptation features are to be implemented, they must not compromise the user's sense of control and should always request permission before proceeding.\\ \cline{2-3}
             & APD 9.i & If any self-adaptation features are to be implemented, they must foster trust between the user and the application by providing a clear, upfront explanation of how and why a self-adaptation was triggered. \\ \cline{2-3}
             & APD 9.j & Users should be given the ability to undo an app’s self-adaptive action intuitively (refer to APD 6.a).\\ \hline
    
    \end{longtable}
}

\begin{figure}
\includegraphics[width=\textwidth]{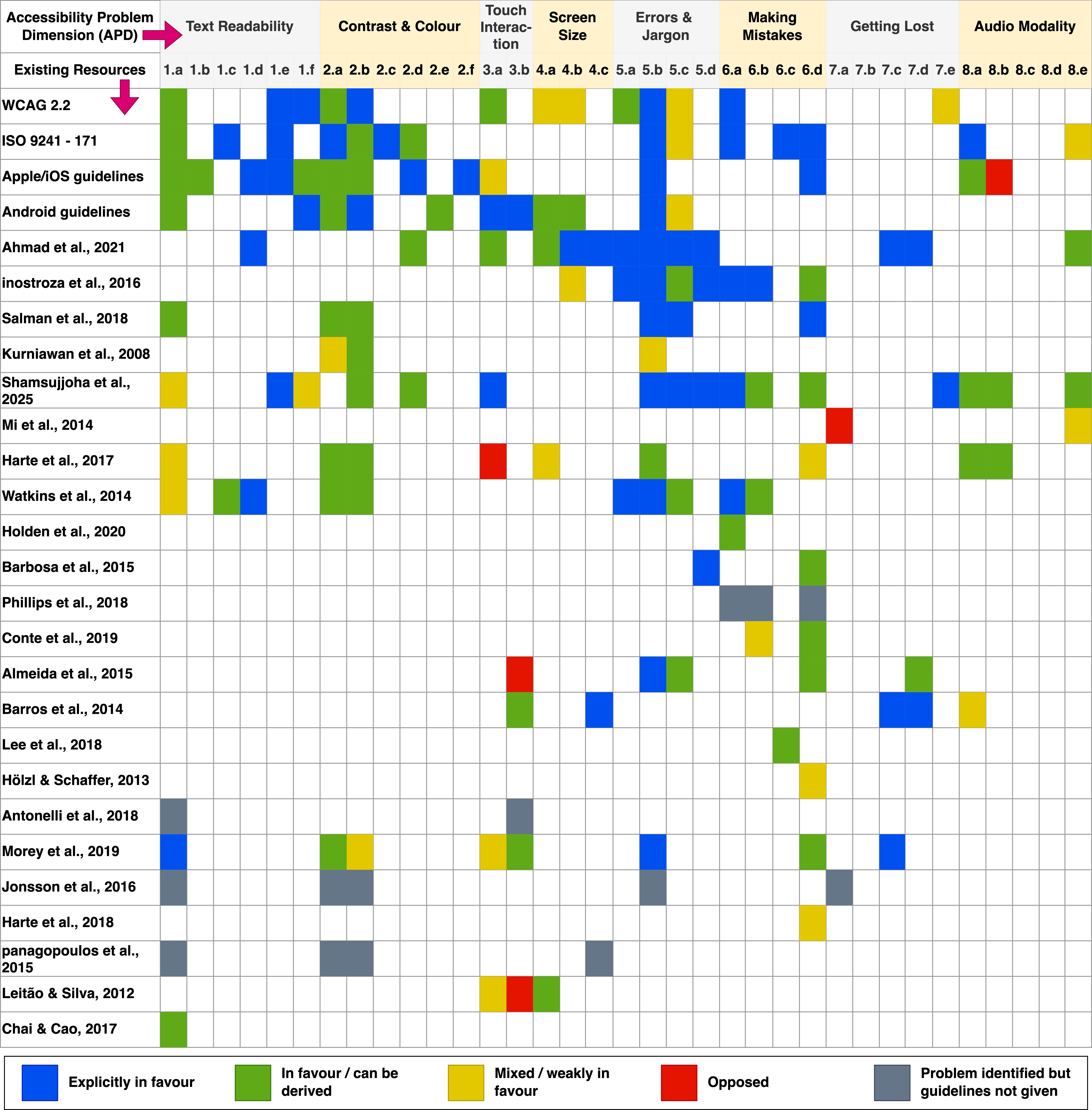}
\caption{A comparison matrix showing how our proposed accessibility guidelines align with existing standards, guidelines, and heuristics related to accessible software design for seniors.}\label{literature_matrix}
\Description{The figure illustrates a literature synthesis matrix that compares the proposed guidelines of this study with existing accessibility standards, platform-specific developer guidelines, and prior research literature, based on the degree of favourability}
\end{figure}

Figure~\ref{literature_matrix} summarises how key related studies and standards correspond with and support our recommendations (excluding the tentative app personalisation recommendations in APD 9). Interestingly, each research study we reviewed proposed different design guidelines, often shaped by variables such as the study’s purpose, the prototypes employed, and the demographic composition of senior participants. The developer-oriented guidelines examined also exhibited notable variation. Similarly, the accessibility challenges identified in our own study could not be fully addressed by relying solely on one or two existing design recommendations. Consequently, we undertook a detailed analysis of each source and synthesised their findings with the empirical insights obtained from our focus group studies. This process enabled us to formulate a coherent set of accessibility recommendations that software practitioners can apply when addressing challenges similar to those identified in our research.

\subsection{Overarching Themes} \label{sec5:themes}

Across all the accessibility aspects and challenges discussed in the previous section, three overarching themes emerged: 
(1) existence of low-tech workarounds among seniors; (2) the diversity of senior accessibility needs; and (3) the personalised nature of senior accessibility needs. 

\subsubsection{\textbf{Low Tech Workarounds}}

Throughout our analysis, we observed instances where seniors circumvented accessibility challenges using practical, low-tech solutions. For example, when screen glare prevented them from reading their screen, they simply moved to a shaded area. In another case, a senior resorted to pen and paper to manually note down their navigation paths within apps.
These examples highlight the resilience and ingenuity of the senior population. However, they also underscore a critical issue: as software practitioners, we have failed to adequately address the accessibility barriers that persist in the software we develop.

\subsubsection{\textbf{Diversity of Senior Accessibility Needs}}

During our longitudinal, multi-phased exploration into senior accessibility needs, we discovered a range of pain points. It is evident that accessibility needs among seniors are highly diverse, significantly affecting their user experience with app and software interfaces. Initially, our exploratory focus groups revealed a broader understanding of the problem space and we sought to address a majority of these issues in the evaluative focus group study. The latter study still continued to generate new insights into accessibility issues faced by seniors. For example, vision impairments can result in varying accessibility needs depending on the specific condition. A senior with macular degeneration may require different accommodations than one with cataracts. Beyond vision-related challenges, other factors, such as cognitive and mobility impairments, cultural differences, and computing literacy levels, further complicate the issue, making accessibility a multifaceted challenge that requires tailored solutions.

\subsubsection{\textbf{Need for App Personalisation}}

During our study, we found that while participants agreed on many accessibility issues and potential solutions, they also expressed contradictory accessibility needs in some areas. One of the most prominent topics of disagreement was error handling in apps.
One group of participants indicated that their first response to an error message was to reboot the app or device. If the same error persisted, they were unwilling to continue using the app and might even uninstall it altogether. Conversely, a more tech-savvy group preferred to receive intuitive error messages with clear resolution steps, demonstrating their willingness to troubleshoot and resolve issues rather than abandon the app.
A similar divergence was observed in UI colour preferences. While a majority of participants stated that they preferred a simple black-and-white interface, one participant reacted positively to the Public Transport Victoria (PTV) app, noting that its subtle use of colour made the interface more visually appealing compared to a purely monochrome design.

These examples highlight a critical consideration in designing for senior users: accessibility needs are not universal. Just as individuals differ in their preferences and limitations, seniors also have varied accessibility requirements, shaped by their personal experiences, technological familiarity, sensory abilities, and other variables.

\section{Limitations}\label{sec5:limitations}

\subsection{Limitations of the Sample}

The primary limitation of our study is the homogeneity of the sample in terms of cultural identity. A majority (65.8\%, 25 out of 38) identified as fully or partially Australian. Among the remaining participants, only four belonged to Asian cultures, while the rest had European cultural backgrounds. Due to this disparity, our findings are most applicable to seniors living in the ‘Western world' (e.g., Australia, New Zealand, and Western Europe). However, we believe that parts of our recommendations can still be generalised to the broader senior community, serving as a valuable foundation for addressing accessibility needs globally.  

Another limitation of the study is that our senior participants demonstrated remarkable confidence in using mobile devices and apps. We suspect Australia's geographical characteristics may influence this, as all participants were recruited from this region. One participant noted that Australians tend to embrace new technologies better compared to other countries in the world. When asked why, we were offered the following explanation: 

\begin{quote}
    \textit{"I think it's also the distance. Oh, in Europe, you're closer. In Asia, you're closer to each other. You're more likely to see family members, whereas a lot of our family members would be all over the place and contacting them, a lot of them like to see each other, especially with lockdowns and stuff."}
\end{quote}  

Thus, the insights and personas derived from our data may be biased towards senior communities with above-average digital literacy and confidence compared to the global senior population. However, it can be argued that our recommendations will become increasingly relevant as current younger generations age and their age-related impairments worsen, given that future seniors will have greater familiarity with technology.

\subsection{Limitations of our Accessibility Recommendations}

The design recommendations presented in this study are not intended to be exhaustive. While they synthesise multiple categories of knowledge and insights, the primary framework used to structure and categorise these recommendations was derived from the themes identified during our focus group studies. As such, the scope of our recommendations is inherently limited to the accessibility challenges that our senior participants chose to discuss during the study. These discussions were further shaped by factors such as the framing of our focus group questions and the specific domain of the prototype application used: retail.

For instance, some participants expressed difficulty in relating to the usefulness of an application focused solely on furniture retail. A prototype situated within a different domain, such as e-banking, might have elicited more relevant insights for some participants; however, this would still neglect other critical domains such as e-health, which may present a distinct set of accessibility challenges. Additionally, practical limitations during the design and development phases required us to prioritise certain accessibility requirements over others. As a result, some cognitively oriented problem themes, such as fear of making mistakes, error handling, and navigation challenges were not fully incorporated into the app demonstration during our evaluative study.

We identify cognitive accessibility as a particularly relevant gap in the current set of recommendations. For instance, one participant shared the accessibility challenges faced by a sibling who had experienced a stroke: an issue that differed significantly from the more commonly discussed themes. Addressing such needs would require the inclusion of seniors with cognitive impairments resulting from sudden trauma in future user studies. Likewise, to adequately address the accessibility requirements of seniors with severe impairments across cognitive, visual, mobility, and psychomotor domains, further empirical investigation is necessary. Expanding our recommendations to encompass these complex scenarios will require dedicated and nuanced studies, given the diversity and severity of impairments involved.

\subsection{Future Work}

A core aspect of our evaluative study was identifying the enablers and barriers seniors face when personalising their apps. These insights will inform the extension of our work with \toolname~\cite{wickramathilaka2025}, ensuring that senior end-users can apply run-time adaptations in their apps. We found that seniors prefer a simple configuration panel for app personalisation and that providing clear, intuitive guidance can help mitigate barriers such as fear of making mistakes and a lack of confidence in their app usage skills.  

The accessibility and personalisation insights and recommendations generated in this study have potential applications beyond mobile interfaces. For instance, many of the recommendations can be readily extended to web-based applications. However, applying these principles to other types of user interfaces, such as those found in emerging technologies, would require further exploration. Technologies such as mixed reality headsets and smart home systems introduce fundamentally different interaction paradigms, necessitating a tailored approach to accessibility design. Smart home environments, in particular, represent a promising area for future research, as they often integrate both visual and auditory interfaces. These systems could play a critical role in supporting the autonomy of seniors, especially in home care settings, thereby amplifying the societal impact of accessible UI design.

Another prevalent theme in our user studies was seniors' concerns about privacy when using apps. As developers, we should focus on enhancing seniors' trust in the apps we create, which could encourage them to experiment with app and device settings to improve accessibility and tailor their experience to their needs. Our approach with \toolname~can be extended to achieve this by allowing developers to embed contextual privacy information as metadata (e.g., terms and conditions related to privacy for an input field) at the app design stage. The tool would then enable adaptations based on a senior user's privacy preferences and the provided privacy metadata. We believe this approach can foster trust and transparency between the app and senior users, ultimately leading to higher satisfaction and engagement.  

\section{Summary}\label{sec6:summary}

Seniors, despite representing a significant and growing segment of society, continue to face substantial challenges when interacting with user interfaces in mobile applications. The primary responsibility for addressing these challenges lies with the software practitioner community, which has yet to establish an optimal approach to eliminating accessibility barriers in app design and development. A core issue is that existing accessibility standards and platform-specific developer resources contain notable gaps when it comes to accommodating seniors as end users. To address this problem space, our objective was to contribute a concrete and actionable set of accessibility design recommendations that software developers can apply in practice.

This study offers several contributions. First, we identified a series of accessibility barriers faced by seniors through a dual-phased focus group study involving 38 participants. These findings were contextualised through the creation of eight fictional personas, which can serve as tools for both practitioners: e.g., in requirements engineering, and researchers aiming to extend this work. Based on the accessibility challenges and corresponding personas, we developed a set of design recommendations aimed at addressing these issues from both senior end-user and developer perspectives. Finally, we explored senior participants’ perceptions of in-app run-time personalisation, identifying preliminary insights into how such features can be provided effectively. These findings will inform the next phase of our work, which seeks to further enhance the accessibility and usability of user interfaces for seniors.

\section{Acknowledgments}

Authors are supported by   Australian   Research   Council(ARC) Laureate Fellowship FL190100035. Haggag is supported by a National Intelligence Post-doctoral Fellowship. Our sincere gratitude goes to the participants who took part in the user study. We also wish to thank participant F1 (anonymised in accordance with ethical considerations), whose support was invaluable in facilitating several of our focus groups at a local chapter of the University of the Third Age.

\bibliographystyle{ACM-Reference-Format}
\bibliography{bibliography}  

\section{Appendices}

\subsection{Accessibility Barriers Encountered - Examples} \label{example codes}

{
    \renewcommand{\arraystretch}{1.3}
    \scriptsize
    \begin{longtable}{|>{\raggedright\arraybackslash}p{2cm} | >{\raggedright\arraybackslash}p{9.5cm} | >{\raggedright\arraybackslash}p{1cm}|}
    \caption{Examples of accessibility barriers discussed by seniors during focus groups.}
    \label{tab:accessibility_barriers} \\
    \hline
    \textbf{Accessibility Barrier} & \textbf{Example Quotes} & \textbf{Focus Group} \\ \hline \hline
    \endfirsthead
    
    \hline
    \textbf{Accessibility Barrier} & \textbf{Example Quotes} & \textbf{Focus Group} \\ \hline \hline
    \endhead
    
    \hline
    \endfoot
    
    \hline
    \endlastfoot
    
                \multirow{5}{*}{\parbox{2cm}{\raggedright APD 1: Text readability}} &
                “Because it's [the example with unreadable text] not good.”\newline\newline
                “Yeah, once it's big enough, I can read it, but the annoying part is moving it backwards and forwards.” & FG-A \\ \cline{2-3}
    
                & “I enlarge it [the text]? I enlarge it so it's comfortable for my own reading.”\newline\newline
                “I can read it [text] with glasses. I'm not sure what it is about the screen, but if I read for a period, so 15 minutes, when I look away from the screen, my vision is quite blurred. So I find that there is a distortion of my vision just for a little while. It needs time then to come back to what I would call the call normal vision”
                & FG-B \\ \cline{2-3}
                
                & “What I was going to say was that a lot of apps these days, when you have to put your address in, it will find it for you. But if your vision is impaired, you can't always tell which one is your address, which is similar.”\newline\newline
                “It's much clearer now, isn't it? the bold and the print, the size of the font.” & FG-C \\ \cline{2-3}
                
                & “It's always the fact that everything is small”\newline\newline
                “When something is really small, I have to go open it over and over to make it bigger, bigger, bigger so I can actually read. Especially, I remember being in the shops, for instance, and I don't have my reading glasses with me, you know, I feel like I'll give up” & FG-D \\ \cline{2-3}
                
                & “It's easier to read, and that's the biggest thing. Like, it's easier to then navigate this than what it looked like it was on the first one”\newline\newline
                “It looks a bit wishy washy to me? It's a bit insignificant. […] It's like, it doesn't stand out strongly. It's wishy washy. […] Not a strong, strong font, strong color, like, it should be more bold, yeah, it needs a stronger contrast, a bolder, a bolder text” & FG-E \\ \hline \hline
    
                \multirow{5}{*}{\parbox{2cm}{\raggedright APD 2.1: Low contrast and use of colour}} &
                “Black on white is the best”\newline\newline
                “Yellow text is a pain, [also] anything in the green. Black text is nice.”
                 & FG-A \\ \cline{2-3}
    
                & “It would put you off reading it. Suppose I can enlarge it and spread it so you can see it more closely, but I don't know how I would change the colour of it. It's already a set colour”\newline\newline
                “If I had an option to go black and white, then I’d go black and white”
                & FG-B \\ \cline{2-3}
                
                & “I'm actually thinking my mother more than me, because she has dry macular degeneration, so she can't see out the center of her eye. She can only see the peripheral so she's she's, for a start, she's got to turn her head every which way to see through the glasses to see the screen, but she can't always see the color, although it's what [redacted] and [redacted] was saying about the contrast between the background and the foreground”\newline\newline
                “I think that you've got them with a gray, pale gray background. I still think that makes it difficult to see. It's not enough contrast between the actual image, little image, and the background for some people. I could see it, but how much longer? We don't know.”
                 & FG-C \\ \cline{2-3}
                
                & “You need to contrast when you get older”\newline\newline
                “Just the darkness or the depth of the writing, the contrast, it's clearer”
                 & FG-D \\ \cline{2-3}
                 
                & “Not a strong, strong font, strong color, like, it should be more bold, yeah, it needs a stronger contrast, a bolder, a bolder text”\newline\newline
                “It's easier to read, and that's the biggest thing. Like, it's easier to then navigate this than what it looked like it was on the first one”
                 & FG-E \\ \hline \hline
    
                \multirow{2}{*}{\parbox{2cm}{\raggedright \rule{0pt}{4ex}APD 2.2: The need for white backgrounds}} &
                “Does it relate to the white white screen as well? Because white screens are always very clear to see on apps”\newline\newline
                “I find white backgrounds just looking generally on the internet, is clearer to see them when there's multi color”
                 & FG-C \\ \cline{2-3}
    
                & “I think that's right. The fact that the background is white”\newline\newline
                “Because you know your eyesight, obviously, is not what it used to be. So you need to have the white background, perhaps, and but you know the contrast, however you achieve that.”
                & FG-D \\ \hline \hline
    
                \multirow{1}{*}{\parbox{2cm}{\raggedright \rule{0pt}{4ex}APD 3: Touch interaction targets being too small}} &
                “When you're older, your eyesight goes I mean, I can see with glasses. But for Mum, it's harder because she it's too small, and she doesn't use a phone except to answer the phone. She only chooses an iPad type”\newline\newline
                “Screens or the boxes are too small” & FG-C \\ \hline \hline
    
                \multirow{5}{*}{\parbox{2cm}{\raggedright APD 4: Limited screen size}} &
                “The thing I think about these two, the templates and the phones compared to computer screens, it's simply the size of the screen. And if, as I would have to increase the text size means the amount of information shown on the screen at one point is very much reduced as compared to a 24-inch computer screen. And for that reason, I find it very difficult, and normally I would go to a computer screen simply because you've got more information on there.”\newline\newline
                “My understanding may be incorrectly, but when we've used looking at the super fund on the phone, It [navigation in phones] a very truncated experience. You don’t have the flexibility. It seems to be written for the limitations of navigation within the small screen that can’t show very much. So, therefore, it seems to be that they [app developers] say: ‘well, what’s the minimum experience we’ve got to give people to make any use at all? And forget about all the extraneous stuff, we’ll leave that to the website” 
                & FG-A \\ \cline{2-3}
    
                & “My wife complains that when that [virtual keyboard] comes up, it can obliterate [her user experience]. Because it fills half the screen. It certainly causes her a lot of frustration. Then what she wanted to look at, she can’t see because the keyboard is there”
                & FG-B \\ \cline{2-3}
                
                & “You can make the text bigger, obviously, but that means the text, especially when it's on largest size. You've got like, two words on the page, you know, so if there was a way to to alleviate that, it would be great.”\newline\newline
                “On my mother's iPads, when she's got the text option maximum, you can never see the except send or the whatever the next, next button, it doesn't appear. You can't get to it.” & FG-C \\ \cline{2-3}
                
                & “When something is really small, I have to go open over and over to make it bigger, bigger, bigger so I can actually read, you know, if I especially mean I remember being in the shops, for instance, and I don't have my reading glasses with me, you know, I feel like I'll give up, you know” 
                & FG-D \\ \cline{2-3}
                
                & “I've made text a little bigger. My screen's only quite small. I've got to be careful with that, because then I lose reading space.”
                & FG-E \\ \hline \hline
    
                \multirow{2}{*}{\parbox{2cm}{\raggedright \rule{0pt}{4ex}APD 5: Error Handling and Jargon}} &
                “One of the most annoying things when you get an error message is they don't actually give you the message. They'll say error 404, what is error 404? So you've gotta go to the internet and type in error 404, and hope the error that they've given you is the same as the one you found on the internet. Why can't they actually tell you what the error is?”\newline\newline
                “It would stop me going back to that website if that happened a second time.” 
                & FG-A \\ \cline{2-3}
    
                & “Turn it on and off again”\newline\newline
                “How’s anyone supposed to know what that means?”\newline\newline
                “Close it and say, I didn't want to know that [error message] anyway....”\newline\newline
                “You also have reference to a number, they might attach a number which helps user to with a very specific explanation of the error. I find it all infuriating. I'd rather have a smiling face, or someone saying, look, maybe we should try this, or something that I find that more seductive than this stuff, this is awful”
                & FG-B \\ \hline \hline
                
                \multirow{5}{*}{\parbox{2cm}{\raggedright APD 6: Fear of making mistakes in apps}} &
                “The difference between a child or a teenager or a young adult doing this, and a senior. The senior is too scared, to touch their button. They're too scared to say, I can actually hit that navigation button because they're too scared that if they do it, something's going to happen to their phone, whereas a person who is used to technology will touch it, but they'll say, I’m still going to hurt my phone. I'm just going to play around with it, and that is the big difference.”\newline\newline
                “But I guess what, what [redacted]'s saying is that if you don't know that, you won't touch it. And that's the biggest and maybe if it was worded as departure times with an arrow, then the person would say, Oh yeah, maybe if I hit that arrow, it would give me different departure times.” 
                & FG-A \\ \cline{2-3}
    
                & “[There are a] lot of schemes that are around. I’m very reluctant to push on anything. I just leave it [error message] or delete it [app]”
                & FG-B \\ \cline{2-3}
                
                & “I think that also the even for me, when you go into accessibility, knowing what each thing does and it doesn't clash with something else, is an issue, you know? I mean, you set it up to do this, you go away, and the next thing, you get a phone call saying it's not working, or I've lost everything you know, or whatever, and then it's another trip back to fix it.” 
                & FG-C \\ \cline{2-3}
                
                & “So for me, fear is a massive one. Yes, I just don't have any confidence in myself to know what I'm doing, to not stuff something up.” \newline\newline
                “Actually, yeah, you don't want to change things that you've become familiar with”
                & FG-D \\ \cline{2-3}
                
                & “I did that trying to change on our little my little notebook, I wanted to get rid of the US English down below, yes, not putting Australian. Well, did that stuff that up right? I thought I'd actually destroyed the little notepad. Was on the phone to my son. He said, You remember how you did it? He said, to reverse it. So, yes, I do do that.”\newline\newline
                “I've done a couple of other things, but I'm very hesitant. I can find my way around that I'm not competent enough comfortable to just go in and do something without clarifying with someone that knows better than what I do.”
                & FG-E \\ \hline \hline
    
                \multirow{2}{*}{\parbox{2cm}{\raggedright \rule{0pt}{4ex}APD 7: Getting lost within apps when navigating}} &

                “I have navigated through, you know, and then found exactly what I want. I don't know how I got there, and to try and find it a second time, you can't, you have no end of trouble.”\newline\newline
                “They [app navigation] are also error-prone in my opinion. That they are error prone to use this. This is quite a detailed navigation, actually, your example. I know that you're doing it for a purpose. But nevertheless, it's easy to be error-prone in these devices.” 
                & FG-A \\ \cline{2-3}
    
                & “Go back to the homepage. Then you know, if I’ve given up looking, I just go to search. I should just probably do that first to save myself the bother of going through all the pages”\newline\newline
                “Bookmarks are really good when you're generally on the internet, see something like that, if that was within, say, Amazon or eBay or whatever, there is no bookmark. It just remembers that you've been there. But in some of them, they don't have a bookmark because you're not actually on the general internet search thing. You're within a program or within a site. So all you can do is you have to remember, by pen and paper where you went and you have to do and it's not always that easy”
                & FG-B \\ \hline
    
                \multirow{4}{*}{\parbox{2cm}{\raggedright APD 8.1: Need for voice input}} 
                & “My wife complains that when that [virtual keyboard] comes up, it can obliterate [her user experience]. Because it fills half the screen. It certainly causes her a lot of frustration. Then what she wanted to look at, she can’t see because the keyboard is there”
                & FG-B \\ \cline{2-3}
                
                & “Because the keyboards are necessarily fairly small given the device size, so therefore, and you know, with dexterity and even without it, I mean, I find that even when I'm doing it, you know, I'll often sort of press something and a number will come up instead of a letter, because I've obviously gone a little bit too fast or something. So I think that would really help older people dexterity and also vision.”\newline\newline
                “Intonation has got a lot to do with it. You're right. Everybody thinks I speak normally, but I know that voice [input] does not interpret my voice today. I think it should know”
                & FG-C \\ \cline{2-3}
                
                & “I think the biggest thing that it is actually addressing is the poor eyesight going back. So then you're already having to deal with a small screen, and having now the bulged letters less on the screen, perhaps a little bit bigger, the fact that you can now use the voice, and you know, the listening means that the eyesight is now being addressed. The problems with eyesight” \newline\newline
                “It could end up being more time consuming if it continues to get it wrong”
                & FG-D \\ \cline{2-3}
                
                & “I think it sometimes makes mistakes.”\newline\newline
                “My husband can't use it. He has a deeper voice. My voice projects, and I can use that. I use that audio, you know, talking [voice input]. I don't know what it's probably called, but I do use that quite a bit for things. But hopefully that aspect of it will be able to enable people with lower registers, or the inability to be able to clearly enunciate”
                & FG-E \\ \hline \hline
    
                \multirow{4}{*}{\parbox{2cm}{\raggedright APD 8.2: Need for screen reading}} &
                “Because you really wouldn't want to… I was in the shopping center this morning. I looked into two conversations in the toilet. For people in the toilet on their phones… [I was] thinking, Oh, how ridiculous! No, put your earphones in. I don't need people hearing what I'm listening to.”
                & FG-A \\ \cline{2-3}
    
                & “I’m actually diverting to audiobooks and things like that if I want to read. I find that I enjoy the audiobooks”\newline\newline
                “I find, on public transport, things like that, I will use that [screen reader], but I will use headphones. Because I've got a great volume [with headphones]”
                & FG-B \\ \cline{2-3}
                
                & “But it's also because we're not used to having that sort of, you know, we haven't been exposed to it, but I think again, because of the eyesight, you know, that is to me, really helpful to look at a screen that is small and then having it say there's a metal table so that you might not see it's metal. You might think it's a color, you know. So to make the description audible to me would be very helpful.” \newline\newline
                “Actually everything, again, because, to me, it is because of eyesight, especially when you're on a small screen. So anything that's then comes out with Audible and obviously we're going to have trouble with our hearing too, you know, over time, I understand that, but because, you know, it might say the price, it might say [something else], you know? Because now, when something is really small, I have to go open over and over to make it bigger, bigger, bigger so I can actually read, you know. Especially mean I remember being in the shops, for instance, and I don't have my reading glasses with me, you know, I feel like I'll give up, you know, because it's just start. But, you know, something audible would be very helpful for me.”
                & FG-D \\ \cline{2-3}
                
                & “Well, you wouldn't, or you could use it on the train, or you could use it in a public space, but you wouldn't use the audio feature, you type it”\newline\newline
                “I wouldn't be using it in a noisy space, because I'm buying furniture, so I wouldn't be out somewhere going furniture. You'd want to be able to concentrate. Yeah, then to be able to concentrate. So you'd be at home.”
                & FG-E \\ \hline \hline
    
                \multirow{2}{*}{\parbox{2cm}{\raggedright \rule{0pt}{4ex}APD 8.3: Validating voice input through screen reading}} &
                “Does that mean we can actually speak into it to say the name, and then you could put press the other button and have it played? Because sometimes I also cut it off. But I used to use the speaking into the messages because it was quicker. But damn, autoCorrect, or whatever they use garbles”
                & FG-C \\ \cline{2-3}
    
                & “Part of the [voice] response, typically, what it does at the end of it, it tells you who you said you were, and then it gives you a chance to go back.”
                & FG-D \\ \hline \hline

                \multirow{3}{*}{\parbox{2cm}{\raggedright \rule{0pt}{4ex}APD 9.1: Personalisation - too much complexity}} &
                “I think a lot of people I know in this [senior] demography would find this dreadful, far too complex and for what? And the strategy today is to ask their [seniors'] grandchildren to do it for them, and so the grandchild sits on the knees, well, what is it you want to do, and leaves everything else out. Getting a senior person who is unfamiliar with this stuff..., it would just be amazing.” \newline\newline
                "If you're talking about just this settings screen, just keeping it as simple [as possible]. Most people aren't going to need all those [adaptation options]. An older person isn't going to worry about changing all this and [so,] keep it very, very simple, the options."\newline\newline
                "Maybe accessibility, when settings are clicked into what's currently there, maybe accessibility, and starting with A) should be on the top, okay, because it's something that will drive what you do with the rest of it." \newline\newline
                "I have to say, I've never even pressed on that to see what it does. So maybe because I haven't needed to, but yeah, if it was at the top, I probably would look at it more readily."\newline\newline
                "I also think that for people my mother's age, which is 90, or, you know, she wouldn't even know what to ask. Oh, okay, so changing the button colour is fairly straightforward, but anything more complex than that, she wouldn't even know how to go about phrasing what words to use."
                & FG-C \\ \cline{2-3}

                & “Like if it [personalisation configuration] was clear and not scary and you didn't think you were sort of signing your life away or giving all your you know, the information to the world..., yeah.” \newline\newline 
                "I think because once the trust is there, and it is in language that we're used to as our generation, the way we learn our generation, then we can familiarize ourselves. And then it becomes just like walking, you know, all of a sudden we know just to press the accessibility button for that particular app"
                & FG-D \\ \cline{2-3}
    
                & “I'm not savvy enough to know what the options are? Yeah, I personally don't think we'd all know what all the possibilities [in app] we could change. I think you have to have some examples”\newline\newline 
                "So I've learned slowly, but I've never done a computer course. So I am hesitant, but I can find my way around, but not necessarily comfortably. And I think that accessibility aspect of you know, we want to change things on apps, like for sales or anything. We might need some guidance about that, and I'm not sure how you would approach that in the apps that you're trying to incorporate." \newline\newline
                "Sometimes I change something and then whoops, that wasn't what I wanted..."
                & FG-E \\ \hline \hline

                \multirow{3}{*}{\parbox{2cm}{\raggedright \rule{0pt}{4ex}APD 9.2: Personalisation - Lack of confidence in technological skills}} &
                "I know a lot of older people are scared of doing it, so they're not. We're familiar with it because we've sort of grown up with it a bit. But a lot of my older friends, they wouldn't use it because of the complexity and the fact that they haven't got the confidence to use it for all those reasons that we've been talking about, apart from anything else"
                & FG-C \\ \cline{2-3}

                & “What if you hit and button and also you cannot no longer.., you know. So for me, fear is a massive one. Yes, I just don't have any confidence in myself to know what I'm doing, to not stuff something up.” 
                & FG-D \\ \cline{2-3}
    
                & 
                "My Ubank app is fabulous. It's geared for any age. And I've made changes in the Ubank app, and it will give you tutorials. Whenever they update things, they say, Hey guys, here it is, you know, and you get a quick snip, snip, snip, and you go, oh, that's a great feature. And if you want to do that, you can do this, and you have this choice. [...] It knows what it can do and it can help you do those things, so long as it gives you, it tells you that you know, would you like to try, try doing it this way, or there's a quicker way or a smarter way."
                & FG-E \\ \hline \hline

                \multirow{3}{*}{\parbox{2cm}{\raggedright \rule{0pt}{4ex}APD 9.4: Personalisation - Reduction of autonomy}} &
                "I find it annoying. Sorry. I would hate it because I'd feel [like] I've lost control."\newline\newline 
                "I would like it, if I understand it. [For example,] if there's something that my grandson has done with it [app/device]: ‘Look quickly, you do it like this.' I think, 'What has he done?' But I just accept it [configuration change]. [When] automatic aspects have [been] changed automatically, I [need to] understand that they've changed automatically, and my grandson hasn't changed it."
                & FG-C \\ \cline{2-3}

                & “I can see how it could be helpful, because it obviously reads your mind as it says, you know, this is what you used to be. Going to go and do it that way. But I think it's a control issue for me, to have control over what I do with an app.” 
                & FG-D \\ \cline{2-3}
    
                & 
                "Not yet! I like a degree of control."\newline\newline 
                "Not without your approval anyway."\newline\newline 
                "You got to control, but don't take it away."\newline\newline
                "All of a sudden, there's this information that artificial intelligence is going to be able to help us buy our furniture, buy this, do all of that, and the general population, me, right? My husband and I, well, we don't know enough about it."
                & FG-E \\ \hline \hline

                \multirow{3}{*}{\parbox{2cm}{\raggedright \rule{0pt}{4ex}APD 9.3: Personalisation - Feeling of Intrusiveness}} &
                "I would like it, if I understand it. [For example,] if there's something that my grandson has done with it [app/device]: ‘Look quickly, you do it like this.' I think, 'What has he done?' But I just accept it [configuration change]. [When] automatic aspects have [been] changed automatically, I [need to] understand that they've changed automatically, and my grandson hasn't changed it."
                & FG-C \\ \cline{2-3}

                & “It feels too intrusive”\newline\newline
                "yes, intrusive somebody else coming and messing with my stuff."
                & FG-D \\ \cline{2-3}
    
                & 
                "As soon as you bring up, AI, you've got some people that are comfortable, and [you] have three quarters of us are going, well, oh, what are we talking about here? I only want to order furniture. Thank you, right? I don't want someone else to and then does this A1 [A one]? AI, right? Does that then go to then read, okay, well, I've looked at Maya for something, and that's useless, and I've looked at that. So all of that information now will just disperse to this."\newline\newline 
                "You say something aloud, something in your house, and all of [a sudden], your Facebook feed feeds you [advertisements]"
                & FG-E \\ \hline \hline
    
    \end{longtable}
}

\subsection{Evidence for Personas} \label{persona evidence}

{
    \renewcommand{\arraystretch}{1.3}
    \scriptsize
    \begin{longtable}{|>{\raggedright\arraybackslash}p{2cm} | >{\raggedright\arraybackslash}p{5cm} | >{\raggedright\arraybackslash}p{5.5cm}|}
    \caption{A summary of evidence for the refined persona corpus based on our focus group studies.}
    \label{tab:persona_evidence} \\
    \hline
    \textbf{Persona} & \textbf{Persona statement} & \textbf{Referenced participant statements} \\ \hline \hline
    \endfirsthead
    
    \hline
    \textbf{Persona} & \textbf{Persona statement} & \textbf{Referenced participant statements} \\ \hline
    \endhead
    
    \hline
    \endfoot
    
    \hline
    \endlastfoot
        \multirow{4}{*}{\parbox{2cm}{\raggedright Persona 01: Carl Jameson (Figure \ref{persona example})}} &
                "Carl has AMD, impairing his central vision and making app content difficult to read. Low-contrast elements make it harder for Carl to distinguish text and UI components, further hindering app usability."
                & 
                “It looks a bit wishy washy to me? It's a bit insignificant. […] It's like, it doesn't stand out strongly. It's wishy washy. […] Not a strong, strong font, strong color, like, it should be more bold, yeah, it needs a stronger contrast, a bolder, a bolder text" -- \textit{[FG-D]} \\ \cline{2-3}

                & "He struggles to accurately tap buttons on the on-screen keyboard and other small UI elements due to his swollen fingers." 
                & "I think if you got larger fingers because my husband has to turn his phone sideways so he doesn’t press too many letters at once \textit{-- [FG-B]}"\\ \cline{2-3}
    
                & 
                "Carl customises his phone’s accessibility settings to improve usability, though these adjustments sometimes cause issues like misaligned UI elements and increased scrolling."
                & "You can make the text bigger, obviously. But that means the text, especially when it's on largest size, You've got like, two words on the page, you know? So if there was a way to to alleviate that, it would be great." -- \textit{[FG-B]}  \\ \cline{2-3}

                & 
                "Carl uses a stylus for more precise touch interactions but occasionally forgets to bring it with him."
                & "I think it’s nice when you can... You know that there’s these pens that you can buy that have that soft points, that make it [virtual keyboard] easier to use" -- \textit{[FG-B]} \\ \hline \hline
                \multirow{4}{*}{\parbox{2cm}{\raggedright Persona 02: Judy Smith (Figure \ref{judy_persona})}} 
                & "She finds it difficult to read small print on the app, even while wearing her glasses." 
                & \multirow{2}{5.5cm}{\raggedright "When something is really small, I have to go open it over and over to make it bigger, bigger, bigger so I can actually read. Especially, I remember being in the shops, for instance, and I don't have my reading glasses with me, you know, I feel like I'll give up." -- [FG-D]} 
                \\ \cline{2-2}
                
                & "She often uses zoom-in gestures to enlarge the text but must navigate the UI with touch gestures back and forth to read the entire content. In some cases, zoom-in gestures are not available in mobile apps." 
                & 
                \\ \cline{2-3}
                
                & 
                "She asked her granddaughter to increase the font size on her phone."
                & "And the strategy today is to ask their [seniors'] grandchildren to do it for them, and so the grandchild sits on the knees, well, what is it you want to do? and leaves everything else out." -- \textit{[FG-C]} \\ \cline{2-3}

                & 
                "She also uses her phone’s landscape mode to read text."
                & "I can turn my phone around so you can actually read it, you know, landscape instead of portrait." -- \textit{[FG-B]} \\ \hline \hline
        Persona 03: Kathryn Davies (Figure \ref{kathryn_persona}) &
                "Kathryn struggles to differentiate certain foreground-background colour combinations due to her clouded and yellow-tinged vision."
                & "Yellow text is a pain, [also] anything in the green." -- \textit{[FG-A]}  \\ 

                & "Using mobile apps outdoors in daylight is very difficult for her due to screen glare." 
                & "You have got your brightness. You can’t really brighten the screen if it’s already brightened to the maximum. But I usually have quite a dull screen. I don't like the brightness of it usually. So if I want to look at something in the sun, I have to go in shade because I can't see it" -- \textit{[FG-B]} \\ \cline{2-3}
    
                & 
                "She enlarges text to improve visibility in low-contrast UI colour combinations."
                & "Suppose I can enlarge it and spread it so you can see it more closely, but I don't know how I would change the colour of it. It's already a set colour" -- \textit{[FG-B]} \\ \cline{2-3}

                & 
                "When using apps outdoors, she moves to a shaded area, though this only provides marginal improvement."
                & "We have to go to the shade, you can’t see the screen that well" -- \textit{[FG-B]} \\ \cline{2-3}

                & 
                "If an app’s contrast is too low, she avoids reading the text and may choose not to use the app at all in the future."
                & "I mean, I've done a  lot of [work] in IT myself. and whenever it came about, planning stuff the web, we were always told you've got to be aware of people's vision and that certain colours people have difficulty seeing, so you've got to prepare your own website in a way that it's going to be generally acceptable, rather than being flashy and to a lot of extent, they were saying it reduced your creativity with regard to how you might want to present your material, but you'll get a wider audience through it being more readable. But that seems to have gone by the board [abandoned] now, because it's just the flashiness which is presented. So, as I said, because of my training, because my awareness, I just don't bother, I'll just get on my computer, if I'm not able to see it" -- \textit{[FG-A]} \\ \hline \hline

        \multirow{4}{*}{\parbox{2cm}{\raggedright Persona 04: Liam O'Connor (Figure \ref{liam_persona})}} &
                "Liam increased his phone’s text size to improve readability, but now he has to scroll more to view content."
                & "I've made text a little bigger. My screen's only quite small. I've got to be careful with that, because then I lose reading space" -- \textit{[FG-E]} \newline \newline
                "Scrolling down the screen is, annoying." -- \textit{[FG-E]}\\ \cline{2-3}

                & "The virtual keyboard takes up half the screen, blocking important content and making navigation frustrating." 
                & "My wife complains that when that [virtual keyboard] comes up, it can obliterate [her user experience]. Because it fills half the screen. It certainly causes her a lot of frustration. Then what she wanted to look at, she can’t see because the keyboard is there" -- \textit{[FG-B]} \\ \cline{2-3}
    
                & 
                "He finds that mobile apps often lack features compared to their web-based versions when accessed on his laptop"
                & "But when we've been used to looking at the super fund on the phone, It [navigation in phones] is a very truncated experience. You don’t have the flexibility. It seems to be written for the limitations of navigation within the small screen that can’t show very much. So, therefore, it seems to be that they [app developers] say: ‘well, what’s the minimum experience we’ve got to give people to make any use at all? And forget about all the extraneous stuff, we’ll leave that to the website'" -- \textit{[FG-B]}  \\ \cline{2-3}

                & 
                "He avoids using mobile apps unless absolutely necessary (e.g., an urgent bank transfer). He prefers to wait until he gets home to complete online tasks on his laptop for a better user experience."
                & "Sometimes it takes a long time to select something and you have to wait a long time until you get to the next page and then you’ve got to go through that and select the items again or another menu and go to something small. So it’s sort of like a starts big and then small and small till you find it down and actually get where you want to go. It'd be lot quicker on the computer" -- \textit{[FG-B]} \\ \hline \hline

        \multirow{4}{*}{\parbox{2cm}{\raggedright Persona 05: Usha Sharma (Figure \ref{usha_persona})}} &
                "Usha struggles with error messages because they often use vague or technical terms (e.g., "404 error"), making them hard to understand."
                & "One of the most annoying things when you get an error message is they don't actually give you the message. They'll say error 404, what is error 404? So you've gotta go to the internet and type in error 404, and hope the error that they've given you is the same as the one you found on the internet. Why can't they actually tell you what the error is?" -- \textit{[FG-A]} \\ \cline{2-3}

                & "When an error occurs, she doesn’t know what to do due to a lack of clear instructions." 
                & "You also have reference to a number, they might attach a number which helps user to with a very specific explanation of the error. I find it all infuriating. I'd rather have a smiling face, or someone saying, look, maybe we should try this, or something that I find that more seductive than this stuff, this is awful" -- \textit{[FG-B]} \\ \cline{2-3}
    
                & 
                "She restarts the app first, hoping the issue fixes itself. If that doesn’t work, she reboots her phone."
                & "You have restart and do all over again to see if it goes through?" -- \textit{[FG-A]} \\ \cline{2-3}

                & 
                "If the error keeps happening, she uninstalls the app."
                & "Why does the customer, who’s paid good money for the application become the regression tester for the designer? That’s what we are doing? That is junk and it [application] should be removed." -- \textit{[FG-B]} \\ \hline \hline

        \multirow{4}{*}{\parbox{2cm}{\raggedright Persona 06: Stefan Bauer (Figure \ref{stefan_persona})}} &
                "Stefan finds it difficult to remember navigation paths, making it hard to replicate a previously completed task."
                & "I have navigated through, you know, and then found exactly what I want. I don't know how I got there, and to try and find it a second time, you can't, you have no end of trouble." -- \textit{[FG-A]} \\ \cline{2-3}

                & "Stefan relies on categorised bookmarks in web apps for easy access, but most mobile apps lack this feature. He finds automated suggestions in apps unhelpful, as they don’t give him the control he prefers." 
                & "With bookmarks, I just save the bookmark. And then every so often, I go in there and can't find it. And then I go and set up categories, and then move the bookmark from there to this one and that one. Yeah, I did that so that they're in categories, so that I can find it again." -- \textit{[FG-A]}\\ \cline{2-3}
    
                & 
                "Stefan uses the search function in apps whenever available."
                & "Go back to the homepage. Then you know, if I’ve given up looking, I just go to search. I should just probably do that first to save myself the bother of going through all the pages" -- \textit{[FG-B]} \\ \cline{2-3}

                & 
                "When search is unavailable, he manually records navigation steps in a physical notebook."
                & "Bookmarks are really good when you're generally on the internet, see something like that, if that was within, say, Amazon or eBay or whatever, there is no bookmark. It just remembers that you've been there. But in some of them, they don't have a bookmark because you're not actually on the general internet search thing. You're within a program or within a site. So all you can do is you have to remember, by pen and paper where you went and you have to do and it's not always that easy." -- \textit{[FG-B]} \\ \hline \hline

        \multirow{4}{*}{\parbox{2cm}{\raggedright Persona 07: Ava Hall (Figure \ref{ava_persona})}} 
                & "Ava knows she can adjust her phone’s settings for better readability but hesitates to do so." 
                & "I've done a couple of other things, but I'm very hesitant. I can find my way around that I'm not competent enough [or] comfortable to just go in and do something without clarifying with someone that knows better than what I do." -- \textit{[FG-E]}
                \\ \cline{2-3}
                
                & "She fears that changing settings might break her phone or that she won’t be able to undo unintended changes." 
                & "What if you hit and button and also you cannot no longer.. you know. So for me, fear is a massive one. Yes, I just don't have any confidence in myself to know what I'm doing, to not stuff something up." -- {[FG-D]}
                \\ \cline{2-3}
                
                & "She asks her son or younger family members for help with settings adjustments." 
                & "I think a lot of people know demography would find this dreadful far too complex and what? And the strategy today is to ask their grandchildren to do it for them, and so the grandchild sits on the knees, well, what is it you want to do, and leaves everything else out. Getting getting a senior person who are unfamiliar with this stuff, it would just be amazing." -- {[FG-C]}
                \\ \cline{2-3}
                
                & "She feels that relying on others reduces her autonomy. She is reluctant to inconvenience others with what she perceives as ‘insignificant’ issues." 
                & "I find it annoying. Sorry. I would hate it because I feel I've lost control." -- [FG-C]\newline\newline
                "I think that also even for me, when you go into accessibility, knowing what each thing does and it doesn't clash with something else, is an issue, you know? I mean, you set it up to do this, you go away, and the next thing, you get a phone call saying it's not working, or I've lost everything you know, or whatever, and then it's another trip back to fix it." -- \textit{[FG-C]}
                \\ \hline \hline

        \multirow{4}{*}{\parbox{2cm}{\raggedright Persona 08: Dorothy Reynolds (Figure \ref{dorothy_persona})}} &
                "Dorothy experiences eye fatigue, making it difficult to read text on her phone for extended periods. She struggles to refocus her vision after reading on her phone for more than 15 minutes."
                & "I can read it <text> with glasses. I'm not sure what it is about the screen, but if I read for a period, so 15 minutes, when I look away from the screen, my vision is quite blurred. So I find that there is a distortion of my vision just for a little while. It needs time then to come back to what I would call the call normal vision" -- [FG-B] \\ \cline{2-3}

                & "She reads content on her phone in small chunks when she has to" 
                & "I read only in small amounts" -- [FG-B] \\ \cline{2-3}
    
                & 
                "She prefers audiobooks over e-books to reduce eye strain."
                & "I’m actually diverting to audiobooks and things like that if I want to read. I find that I enjoy the audiobooks" -- [FG-B] \\ \cline{2-3}

                & 
                "She uses the screen reader functionality but disables it when leaving home, due to privacy concerns or situational inappropriateness."
                & "I don't like people that have anything on loudspeaker in the public so I wouldn't be listening to shopping." -- [FG-D] \newline\newline
                "If I was going to do something where, particularly if it was going [on] my phone, it's a bit old and and the actual volume isn't really good, so quite often I'll need to press speaker for it to actually come back to me. So if I had speaker, everybody would know everything that might.. address, personal details and even [things I] just want to order." -- [FG-C]
                \\ \hline
    \end{longtable}
}

\subsection{Accessibility Guidelines Analysis: Synthesis Matrix} \label{Accessibility Guidelines Analysis}

\noindent Link: \url{https://github.com/ShavindraWic/ToSEM-focus-group-paper-replication-package.git} 

\end{document}